\documentclass[pdflatex,sn-mathphys-num]{sn-jnl}

\usepackage{graphicx}%
\usepackage{multirow}%
\usepackage{amsmath,amssymb,amsfonts}%
\usepackage{amsthm}%
\usepackage{mathrsfs}%
\usepackage[title]{appendix}%
\usepackage{xcolor}%
\usepackage{textcomp}%
\usepackage{manyfoot}%
\usepackage{booktabs}%
\usepackage{algorithm}%
\usepackage{algorithmicx}%
\usepackage{algpseudocode}%
\usepackage{listings}%
\usepackage{hyperref}
\usepackage[acronym]{glossaries}
\usepackage[version=4]{mhchem}

\theoremstyle{thmstyleone}

\theoremstyle{thmstyletwo}

\theoremstyle{thmstylethree}
\newtheorem{definition}{Definition}

\theoremstyle{definition}
\newtheorem{recommendation}{Recommendation}

\makeglossaries

\newacronym{randd}{R\&D}{Research and Development}

\newacronym{sdg}{SDG}{Sustainable Development Goal}
\newacronym{ghg}{GHG}{Greenhouse Gas}
\newacronym{co2}{\ce{CO2}}{Carbon Dioxide}
\newacronym{co2e}{\ce{CO2}e}{Carbon Dioxide Equivalent}
\newacronym{ree}{REE}{Rare Earth Elements}
\newacronym{gwp}{GWP}{Global Warming Potential}
\newacronym{gwi}{GWI}{Global Warming Impact}
\newacronym{lca}{LCA}{Life Cycle Assessment}
\newacronym{lcia}{LCIA}{Life Cycle Impact Assessment}
\newacronym{pfas}{PFAS}{Per- and PolyFluoroAlkyl Substances}
\newacronym{iso}{ISO}{International Standards Organisation}
\newacronym{trl}{TRL}{Technology Readiness Level}
\newacronym{cop}{COP}{Co-efficient of Performance}
\newacronym{fair}{FAIR}{Findable, Accessible, Interoperable and Reusable}
\newacronym{ci/cd}{CI/CD}{Continuous Integration and Continuous Delivery}
\newacronym{ccs}{CCS}{Carbon Capture and Storage}
\newacronym{ges}{GES}{Gaz à Effet de Serre}
\newacronym{hpc}{HPC}{High Performance Computing}
\newacronym{capex}{CapEx}{Capital Expenditure}
\newacronym{opex}{OpEx}{Operating Expenses}
\newacronym{breeam}{BREEAM}{Building Research Establishment Environmental Assessment Method}

\newacronym{un}{UN}{United Nations}
\newacronym{stfc}{STFC}{Science and Technology Facilities Council}
\newacronym{ukri}{UKRI}{United Kingdom Research and Innovation}
\newacronym{dl}{DL}{Daresbury Laboratory}
\newacronym{ruedi}{RUEDI}{Relativistic Ultrafast Electron Diffraction \& Imaging}
\newacronym{ilc}{ILC}{International Linear Collider}
\newacronym{clic}{CLIC}{Compact Linear Collider}
\newacronym{cern}{CERN}{European Organization for Nuclear Research}
\newacronym{fcc}{FCC}{Future Circular Collider}
\newacronym{spade}{SPADE}{Sustainability Principles and Advice for Design and Engineering}
\newacronym{dpf}{DPF}{Division of Particles and Fields}
\newacronym{fwf}{FWF}{Austrian Science Fund}
\newacronym{nwo}{NWO}{Dutch Research Council}
\newacronym{embo}{EMBL}{European Molecular Biology Organization}
\newacronym{embl}{EMBL}{European Molecular Biology Laboratory}
\newacronym{fnp}{FNP}{Foundation for Polish Science}
\newacronym{anr}{ANR}{French National Research Agency} 
\newacronym{dfg}{DFG}{German Research Foundation}
\newacronym{gln}{GLN}{Green Labs Netherlands}
\newacronym{ibec}{IBEC}{Institute for Bioengineering of Catalonia}
\newacronym{mrc}{MRC}{Medical Research Council}
\newacronym{cnrs}{CNRS}{National Centre for Scientific Research}
\newacronym{aps}{APS}{American Physical Society}
\newacronym{desy}{DESY}{Deutsches Elektronen-Synchrotron}
\newacronym{esrf}{ESRF}{European Synchrotron Radiation Facility}
\newacronym{esppu}{ESPPU}{European Strategy for Particle Physics Update}
\newacronym{kit}{KIT}{Karlsruhe Institute of Technology}
\newacronym{swifthep}{SWIFT-HEP}{SoftWare InFrasTructure for High Energy Physics}
\newacronym{bnl}{BNL}{Brookhaven National Lab}
\newacronym{ess}{ESS}{European Spallation Source}
\newacronym{hllhc}{HL-LHC}{High-Luminosity Large Hadron Collider}
\newacronym{susfecit}{SUSFECIT}{Sustainable Federated Compute Infrastructures}
\newacronym{rf2.0}{RF2.0}{Research Facilities 2.0}
\newacronym{green disc}{Green DiSC}{Green Digital Sustainability Certification}
\newacronym{grand}{GRAND}{Giant Array for Neutrino Detection}
\newacronym{dphep}{DPHEP}{Data Preservation in High Energy Physics}
\newacronym{ifast}{I.FAST}{Innovation Fostering in Accelerator Science and Technology}
\newacronym{alice}{ALICE}{A Large Ion Collider Experiment}
\newacronym{erum}{ErUM}{Erforschung von Universum and Materie}
\newacronym{hecap+}{HECAP+}{High Energy Physics, Cosmology, Astroparticle Physics and related disciplines}
\newacronym{icfa}{ICFA}{International Committee for Future Accelerators}
\newacronym{isas}{iSAS}{Innovate for Sustainable Accelerating Systems}
\newacronym{flexrican}{FlexRICAN}{Flexibility in RIs for global CArbon Neutrality}
\newacronym{eajade}{EAJADE}{Europe-America-Japan Accelerator Development Exchange Programme}
\newacronym{esfri}{ESFRI}{European Strategy Forum on Research Infrastructures}
\newacronym{aries}{ARIES}{Accelerator Research and Innovation for European Science and Society}
\newacronym{eucard-2}{EuCARD-2}{Enhanced European Coordination for Accelerator Research \& Development}
\newacronym{guilt}{GUILT}{Green Usage Impact Logging Tool}
\newacronym{essri}{ESSRI}{Energy for Sustainable Science at Research Infrastructures}
\newacronym{hep}{HEP}{High Energy Physics}
\newacronym{ipac}{IPAC}{International Particle Accelerator Conferenc}
\newacronym{ichep}{ICHEP}{International Conference on High Energy Physics}
\newacronym{lhc}{LHC}{Large Hadron Collider}
\newacronym{wlcg}{WLCG}{Worldwide LHC Computing Grid}
\newacronym{sc4rc}{SC4RC}{Sustainability Conference for Responsible Research Computing}
\newacronym{seri}{SERI}{Swiss State Secretariat for Education Research and Innovation}
\newacronym{jai}{JAI}{John Adams Institute for Accelerator Science}
\newacronym{cbm}{CBM}{Compressed Baryonic Matter}
\newacronym{esablim}{ESABLIM}{Energy SAving Beam Line Magnets}
\newacronym{cas}{CAS}{CERN Accelerator School}

\newacronym{sc}{SC}{Superconducting}
\newacronym{nc}{NC}{Normal Conducting}
\newacronym{hts}{HTS}{High Temperature Superconducting}
\newacronym{rf}{RF}{Radio Frequency}
\newacronym{srf}{SRF}{Superconducting Radio Frequency}
\newacronym{pmu}{PMU}{Power Management Unit}
\newacronym{zepto}{ZEPTO}{Zero Power Tuneable Optics}
\newacronym{hepto}{HEPTO}{Hybrid Electromagnet-Permanent magnet Tuneable Optics}
\newacronym{frt}{FRT}{Fast Reactive Tuners}
\newacronym{pv}{PV}{Photovoltaics}
\newacronym{smr}{SMR}{Small Modular Reactors}
\newacronym{hvac}{HVAC}{Heating, Ventilation, and Air Conditioning}
\newacronym{mw}{MW}{Megawatts}
\newacronym{cpu}{CPU}{Central Processing Unit}
\newacronym{gpu}{GPU}{Graphics Processing Unit}
\newacronym{pue}{PUE}{Power Usage Efficiency}
\newacronym{epn}{EPN}{Event Processing Nodes}
\newacronym{eem}{EEM}{Efficient Energy Management}
\newacronym{it}{IT}{Information Technology}
\newacronym{fss}{FSS}{Fair Share Scheduling}
\newacronym{iot}{IOT}{Inductive Output Tube}
\newacronym{esg}{ESG}{Environmental, Social, Governance}
\newacronym{mcda}{MCDA}{Multi-Criteria Decision Analysis}
\newacronym{cba}{CBA}{Cost-Benefit Analysis}

\begin{document}

\title[High-level environmental sustainability guidelines for large accelerator facilities]{High-level environmental sustainability guidelines for large accelerator facilities}
\subtitle{Living document: Version 2}

\author*[1]{\fnm{Hannah} \sur{Wakeling}}\email{hannah.wakeling@physics.ox.ac.uk}%\orcidlink{0000-0003-4606-7895}
\author[1]{\fnm{Philip N.} \sur{Burrows}}%\orcidlink{0000-0002-4518-526X}}
\author[2,3]{\fnm{Jim} \sur{Clarke}}
\author[4]{\fnm{Jo} \sur{Colwell}}
\author[5]{\fnm{Niko} \sur{Neufeld}}%\orcidlink{0000-0003-2298-0102}
\author[2,3]{\fnm{Ben} \sur{Shepherd}}%\orcidlink{0000-0001-9784-3781}
\author[6]{\fnm{Dwayne} \sur{Spiteri}}%\orcidlink{0000-0002-9226-2539}
\author[7]{\fnm{John} \sur{Thomason}}

\affil*[1]{\orgdiv{John Adams Institute}, \orgname{University of Oxford}, \orgaddress{\street{Keble Road}, \city{Oxford}, \postcode{OX1 3RH}, \state{Oxfordshire}, \country{UK}}}
\affil[2]{\orgdiv{ASTeC}, \orgname{UKRI}, \orgaddress{\street{STFC Daresbury Laboratory}, \city{Daresbury}, \postcode{WA4 4AD}, \state{Warrington}, \country{UK}}}
\affil[3]{\orgdiv{The Cockcroft Institute}, \orgname{Sci-Tech Daresbury}, \orgaddress{\street{Keckwick Lane}, \city{Daresbury}, \postcode{WA4 4AD}, \state{Warrington}, \country{UK}}}
\affil[4]{\orgdiv{STFC}, \orgname{UKRI}, \orgaddress{\city{Didcot}, \postcode{OX11 0QX}, \state{Oxfordshire}, \country{UK}}}
\affil[5]{\orgdiv{EP Department}, \orgname{CERN}, \orgaddress{\street{Esplanade des Particules 1}, \city{Meyrin}, \postcode{1217}, \state{Geneva}, \country{Switzerland}}}
\affil[5]{\orgname{Deutsches Elektronen-Synchrotron DESY}, \orgaddress{\street{Notkestrasse 85}, \postcode{22607 Hamburg}, \country{Germany}}}
\affil[7]{\orgdiv{ISIS Neutron and Muon Source}, \orgname{STFC Rutherford Appleton Laboratory}, \orgaddress{\city{Didcot}, \postcode{OX11 0QX}, \state{Oxfordshire}, \country{UK}}}


\abstract{The proposed construction of new particle accelerator-based facilities in the coming decades -- and upgrades to existing facilities -- provides the unique opportunity to embed innovative environmental impact reduction techniques into their design. 
This living document provides high-level guidelines to improve environmental sustainability in the planning, construction, operational and decommissioning stages of large accelerator facilities. A collection of various resources is provided, with examples of some existing and suggested practices.}




\maketitle

\begin{center}
\centering{\textbf{Please consider reading this document via the least environmentally impactful method available to you.}}
\end{center}

\section{Introduction}

Particle accelerator-based facilities have the ability to allow humanity to investigate the fundamental building blocks of the universe, revolutionise modern medicine, understand and create materials with incredible properties, fundamentally expand the boundaries of human knowledge, and much more. 
Whilst large accelerator facilities\footnote{Large accelerator facilities are defined here as facilities that accelerate particles to high energies, have substantial physical structures and consist of multiple subsystems.} have been shown to have a positive impact on society through their scientific output, and economic and environmental benefits, they also negatively impact the environment in a way that is directly related to their size and/or resource consumption. 

As the scientific reach of these facilities grows, often the environmental impact to run them also grows.
Yet, the climate crisis calls for increasingly urgent action towards the reduction and mitigation of probable effects. 
The European Strategy Group for Particle Physics (2020) recommended that ``a detailed plan for the minimisation of environmental impact and for the saving and re-use of energy should be part of the approval process for any major project''\cite{ESPPU-2020}, implying that sustainability is intrinsic to the future of accelerator facilities. Therefore, it is essential that the entire environmental impact of accelerator facilities is understood and minimised. 

The field of particle accelerator environmental sustainability, whilst a relatively young one, already boasts a wide range of studies, efforts and results. This living document provides high-level guidelines, targeted specifically at large accelerator facilities based on these efforts, to aid in the pursuit of improved environmental sustainability. For these facilities to truly incorporate and contribute towards environmental sustainability, a collection of resources is provided to highlight some suggested practices and case studies (where they exist). 

This section further introduces sustainability as it pertains to large accelerator facilities.
Section~\ref{sec:sust-large-acc} discusses the general ways in which an accelerator facility could refocus towards environmental sustainability, and provides accelerator facility specific recommendations for areas to investigate for potential environmental impact reduction.
Finally, Section~\ref{sec:resources} collates additional resources that may be beneficial in knowledge transfer, and in the evaluation and reduction of environmental impact at facilities.

%%%%%%%%%%%%%%%%%%%%%%%%%%%%%%%%%%%%%%%%
\subsection{Sustainability and key definitions \label{sec:sustainability}}

\begin{figure}[h]
    \centering
    \makebox[0.5\textwidth][c]{\includegraphics[width=0.8\textwidth]{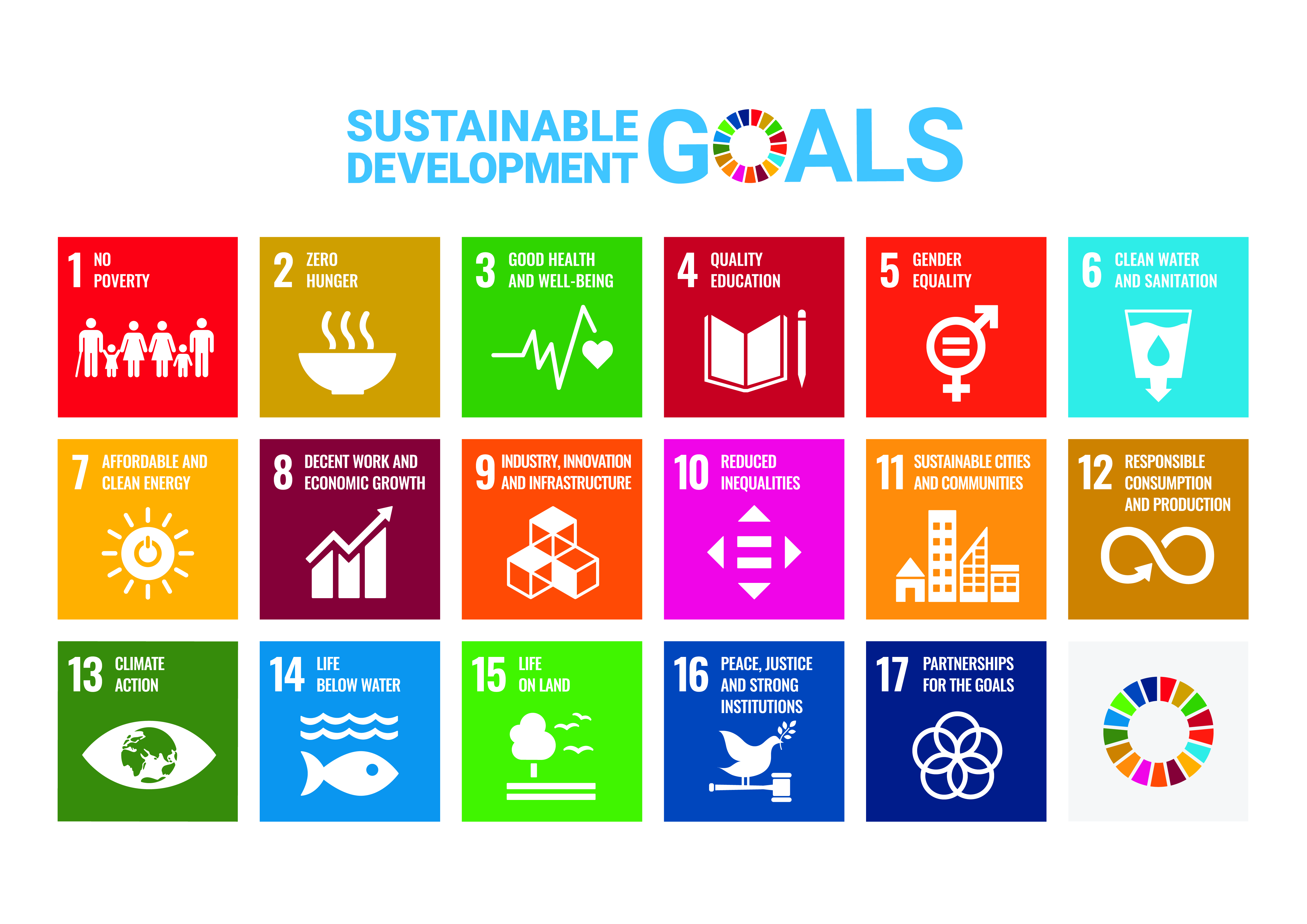}}%
    \caption{The 17 \acrshort{un} \acrshort{sdg}s~\cite{SDGs}, explicitly: 1. No Poverty; 2. Zero Hunger; 3. Good Health and Well-being; 4. Quality Education; 5. Gender Equality; 6. Clean Water and Sanitation; 7. Affordable and Clean Energy; 8. Decent Work and Economic Growth; 9. Industry, Innovation and Infrastructure; 10. Reduced Inequality; 11. Sustainable Cities and Communities; 12. Responsible Consumption and Production; 13. Climate Action; 14. Life Below Water; 15. Life on Land; 16. Peace, Justice and Strong Institutions; and 17. Partnerships to achieve the Goals. The content of this publication has not been approved by the United Nations and does not reflect the views of the United Nations or its officials or Member States. (\href{https://www.un.org/sustainabledevelopment}{Link to UN SDGs website})}
    \label{fig:sdgs}
\end{figure}

\begin{definition}\textit{Sustainability}, particularly the Brundtland definition of sustainability, is defined as meeting the needs of the current generation without impact on future generations~\cite{BrundtlandSustainability}. 
It is widely considered to have three main pillars: economic, social and environmental sustainability. This document focuses on environmental sustainability but encourages further consideration of the other -- equally important -- pillars. 
In particular, the impact of accelerator facilities on the 17 United Nations (\acrshort{un}) Sustainable Development Goals (\acrshort{sdg}s) should be considered (Figure~\ref{fig:sdgs}). Of these 17 UN SDGs, the most relevant ones to (large) accelerator facilities are:
\begin{itemize}
    \item The high consumption of energy of large accelerator facilities (up to the order of Terawatt-hours per year (TWh/yr)) factors directly into the 7\textsuperscript{th} SDG of affordable and clean energy for all with its immediate impact on a global level. 
    \item The radioactive and non-radioactive waste produced by large accelerator facilities must be contained with e.g. sufficient shielding and procedures to prevent e.g. contaminants and heat impacting the ecosphere. Thus the 6\textsuperscript{th}, 14\textsuperscript{th}, and 15\textsuperscript{th} SDGs must be considered.
    \item The influence and direct contributions that large accelerator facilities have on innovation, industry and infrastructure make them key to the 9\textsuperscript{th} SDG.
    \item The (method of) resource consumption of large accelerator facilities impacts the 12\textsuperscript{th} SDG.
    
\end{itemize}
\end{definition}

True environmental sustainability exhibits circularity, thus resource (over)consumption without counteraction is, by definition, unsustainable. This suggests that many systems cannot be truly environmentally sustainable, including accelerator facilities. However, they can reduce their environmental impacts and \textit{improve} their sustainability. Actions can and should be taken to reduce and mitigate resource consumption and material use such that maximal sustainability and minimal environmental impact are achieved. 

\begin{definition}\textit{Systems thinking} is the concept of considering a problem or subject as a whole, rather than trying to consider it in parts~\cite{Systems_Thinking}. It has been gaining traction in relation to tackling the climate crisis due to the potential for system wide changes required to transition to a more sustainable world. As the climate crisis is a `wicked problem'\footnote{A wicked problem, first coined by Rittel and Webber in 1973~\cite{wicked-problem}, is a unique and ever-evolving problem that has no definitive solution, and has interdependency and value conflicts. In contrast to the wicked problem is the `tame' or `benign' problem, in which a mission and whether the mission has been solved is clear. Tame problems are regularly encountered by scientists and engineers and could include examples such as the problem of reducing beam loss to $<$0.001\%, or the problem of developing a 50\,T superconducting magnet.}, with issues impacting and impacted by environment, society, economics, etc., a systems approach using a lifecycle perspective may be able to consider and address this multifaceted problem in a holistic manner.
\end{definition}

Systems thinking is of particular importance to the field of accelerator physics due to the often complex and interdependent systems within it. In the quest for more sustainable research, full lifecycles should be examined to ensure unintentional and unexpected negative consequences are identified and avoided. This will need to be done through robust impact assessments, such as \acrlong{lca}s (\acrshort{lca})~\cite{iso14040,iso14044}, performed on a case-by-case basis, as discussed in Section~\ref{sec:evaluation}. In addition, systems thinking should be tackled through internal and external collaboration, and interdisciplinary knowledge sharing and training (see Section~\ref{sec:skillsets}).

Consideration of environmental sustainability incorporates all types of environmental impacts, which can be measured through various environmental impact factors. These include factors such as global and regional impacts, e.g. global warming potential through increasing Greenhouse Gases (\acrshort{ghg}s) in the atmosphere; resource use, e.g. energy, water and the depletion of mineral and fossil resources; and human and ecological health, e.g. ecotoxicity and fine particulate matter formation. 

When considering and evaluating environmental impact, all impact factors have importance, and should be taken into account. Often, however, a conscious choice is made to focus more on reducing the global warming potential factor, i.e. GHG or carbon impact, in order to tackle the high priority wicked problem.

\begin{definition}\textit{Net zero} is a Carbon Dioxide (\acrshort{co2}) framework towards climate action and sustainability: ``CO$_2$-induced warming halts when net anthropogenic CO$_2$ emissions halt [...] with the level of warming determined by cumulative net emissions to that point''~\cite{netzero}. An organisation or country that can reach net zero emissions would be carbon neutral, i.e. their overall carbon emissions are zero. 
\end{definition}

\begin{definition}\textit{Embodied carbon} refers to the total GHG emissions generated in the construction of a building or other asset. It includes the Carbon Dioxide Equivalent (\acrshort{co2e}) emissions from material extraction to end-of-life disposal. It does not include operational resource consumption.
\end{definition}

Defined by the GHG Protocol in 2001~\cite{GHGprotocol2001}, an organisation's CO$_2$e emissions can be categorised into 3 scopes.

\begin{definition}\textit{Scope 1}
%[leftmargin=!,labelwidth=\widthof{\bfseries Scope 1:}]
contains the direct CO$_2$e emissions of an organisation from sources that the organisation owns or controls, e.g. direct fuel emissions (diesel/petrol) and other CO$_2$e gases, like methane and fluorinated gases.
\end{definition}
\begin{definition}\textit{Scope 2}
contains the indirect CO$_2$e emissions of an organisation, e.g. the electricity purchased to run a facility.
\end{definition}
\begin{definition}\textit{Scope 3}
contains the indirect CO$_2$e emissions of an organisation's value chain, e.g. suppliers of extracted metals. This is broken down into upstream (emissions related to the reporter's suppliers) and downstream (emissions related to the reporter's customers) emissions. In the case of an accelerator facility, an example of upstream emissions could be business travel or waste from operations, and an example of downstream emissions could be leased assets or use of sold products.
\end{definition}

Carbon neutrality specification criteria, such as the \acrshort{iso} 14068~\cite{iso14068} criteria, should be used alongside the GHG scopes to determine compliance with organisational policies. 

Reaching net zero does not equate to being environmentally friendly, just `carbon neutral'\footnote{The term `carbon neutral' includes CO$_2$e GHGs.}. It is wise to mitigate other environmental impacts such as those on biodiversity. Furthermore, net zero is not the end-goal; humanity must continue removing GHGs from the atmosphere after reaching neutrality\footnote{The `safe' level of CO$_2$ in the atmosphere is 350 parts per million (ppm). Preliminary measurements by NOAA in December 2025 measured CO$_2$ levels to be 427.49\,ppm~\cite{NOAA_CO2}.} and continue to embrace environmental stewardship.

%%%%%%%%%%%%%%%%%%%%%%%%%%%%%%%%%%%%%%%%
\subsection{Organisational environmental sustainability policies \label{sec:policy}}

Accelerator facility sustainability goals are intrinsically linked to the legislation of national and international institutions, such as those of intergovernmental organisations (such as the \acrshort{un} \acrshort{sdg}s, or those of a country's leadership (e.g. the United Kingdom's Greening Government Commitments~\cite{UKggcs}). Many of these goals, policies and guidelines have now been accepted into -- and sometimes further strengthened within -- organisational legislation such as that at the Science and Technology Facilities Council (\acrshort{stfc})~\cite{STFCpolicy} or the United Kingdom Research \& Innovation (\acrshort{ukri}) sustainability policies~\cite{UKRIpolicy}.

Knowledge of applicable policies is essential during the planning, construction, operational lifetime and decommissioning of large accelerator facilities. 
The contribution of the facilities to meeting respective organisational goals is paramount. 
It is also recommended that the facility does not just strive to meet policy but goes above and beyond. 
Efforts should be made to pre-empt the need for increased sustainability. 
For example, it should be expected that environmental sustainability will need to be an intrinsic part of any proposal, including considering the impact of an experiment's scientific output and the \acrshort{randd} required. This could manifest itself as environmental assessments, fossil fuel divestment, and emissions reduction goals and/or budgets required for funding bids.

%%%%%%%%%%%%%%%%%%%%%%%%%%%%%%%%%%%%%%%%
\subsubsection*{UN policy \label{sec:UNpolicy}}

The United Nations Framework Convention on Climate Change~\cite{UNFCCC} was founded in 1994. Later, the agreement termed the Paris Agreement~\cite{ParisAgreement} was created, in which 196 parties committed to reduce their \acrshort{ghg} emission levels so as to keep the global rise in temperature to well below 2$^\circ\,$C above pre-industrial levels.
This implies that accelerator facilities built and operated by members of these parties should be realised with environmental responsibility.

%%%%%%%%%%%%%%%%%%%%%%%%%%%%%%%%%%%%%%%%%%%%%%%%%%%%%%%%%%%%%%
\subsubsection*{European Strategy for Particle Physics}

The European Strategy for Particle Physics, first published in 2006, has since been updated in 2013 and 2020. The third update -- the European Strategy for Particle Physics Update (\acrshort{esppu}) -- was launched by the \acrshort{cern} Council in June 2024 and will be published in 2026. The recommendations from the European Strategy Group have been made to the CERN Council and were published in December 2025~\cite{ESPPU-2026}. 

The first edition that explicitly included prominent mention of environmental sustainability was the 2020 update, in which the continued study and minimisation of environmental impact of particle physics is recommended~\cite{ESPPU-2020}. Community submissions towards the European Strategy for Particle Physics updates have increasingly included various levels of calls for environmental sustainability consideration~\cite{ESPPU-2020-ECR, ESPPU-2026-ECR, independent-ESPPU, LDG-ESPPU}. 

The recommendations from the European Strategy Group towards the CERN Council for the 2026 update includes the following on environmental sustainability~\cite{ESPPU-2026}:
\begin{itemize}
    \item ``For new proposed projects, a detailed life cycle assessment should be carried out at each stage, from concept to design and implementation, in order to quantify and minimise environmental  impact.''
    \item ``The particle physics community should continue and intensify its efforts to develop and adopt sustainable solutions.''
    \item ``An effective balance between in-person and online meetings should be considered in order to mitigate the environmental impact of carbon-intensive travel.''
\end{itemize}

%%%%%%%%%%%%%%%%%%%%%%%%%%%%%%%%%%%%%%%%%%%%%%%%%%%%%%%%%%%%%%
\subsubsection*{Snowmass}
The Division of Particles and Fields (\acrshort{dpf}) of the American Physical Society (\acrshort{aps}) organised the Particle Physics Community Planning Exercise (a.k.a. “Snowmass”) in 2021. The report of the Topical Group on Environmental and Societal Impacts of Particle Physics for Snowmass~\cite{SNOWMASS-env} provides 8 recommendations relating to impacts on climate, notably highlighting the importance of evaluation and reporting, everyone working to reduce impacts, and communication. Following this, the overall DPF summary document states, in relation to environmental sustainability: ``Finally, HEP must take greater responsibility for its impacts on climate change by addressing and mitigating these impacts through DOE project policies and individual community member actions''~\cite{Snowmass2021summary}.

\subsubsection*{The Heidelberg Agreement}
The Heidelberg Agreement~\cite{HeidelbergAgreement} includes 9 recommendations on environmental sustainability in research funding. 
It highlights the importance of embedding sustainability into research funding and gives funders concrete recommendations on how to do so.
It was published in October 2024 and endorsed by the Austrian Science Fund (\acrshort{fwf}), Dutch Research Council (\acrshort{nwo}),  European Molecular Biology Organization (\acrshort{embo}), European Molecular Biology Laboratory (\acrshort{embl}), Foundation for Polish Science (\acrshort{fnp}), French National Research Agency (\acrshort{anr}), German Research Foundation (\acrshort{dfg}), Green Algorithms Initiative, Green Labs Netherlands (\acrshort{gln}), Institute for Bioengineering of Catalonia (\acrshort{ibec}), Medical Research Council (\acrshort{mrc}), National Centre for Scientific Research (\acrshort{cnrs}), UK Research \& Innovation (\acrshort{ukri}) and The Wellcome Trust.

%%%%%%%%%%%%%%%%%%%%%%%%%%%%%%%%%%%%%%%%%%%%%%%%%%%%%%%%%%
\clearpage

{\color{gray}\hrule}
\section{Refocusing large accelerator facilities towards environmental sustainability \label{sec:sust-large-acc}}
{\color{gray}\hrule}
\vspace{1.5em}

%%%%%%%%%%%%%%%%%%%%%%%%%%%%%%%%%%%%%%%%%

This section outlines key areas for investigating potential reductions in the environmental impact of a large accelerator facility. A non-exhaustive list of suggestions for improvement of environmental sustainability is provided, with reference to resources where available.

As previously stated, this is a living document. Constructive feedback and suggestions are welcomed by the authors. For each suggestion, due diligence and consideration are necessary, on a case-by-case basis, to ensure that any actions taken would indeed reduce the accelerator facility's environmental impact.

There is substantial overlap between the following subsections that should be acknowledged. Efforts have been made to explicitly point out less obvious overlap where applicable, but in the quest for a succinct document, some more obvious links are not stated. One key takeaway, discussed in Section~\ref{sec:prioritisation}, is that future operational and decommissioning environmental reductions need to be enabled within the planning, design and construction stage of the facility.

\begin{recommendation}\textbf{Sustainability efforts be ranked in preference and impact by the 7 R's: Rethink, Refuse, Reduce, Reuse, Repair, Repurpose, and Recycle.} Priority should be given to preventing impacts. 
\end{recommendation}

%%%%%%%%%%%%%%%%%%%%%%%%%%%%%%%%%%%%%%%%
\subsection{Culture change \label{sec:culture}}

The action with the greatest potential for impact on environmental sustainability is culture change~\cite{ClimateChangeUNESCO, CultureChangeNZ}, yet it is one of the hardest to implement and most difficult for which to measure success. 
Culture change in relation to sustainability allows for the improvement of awareness, attitude and actions towards reducing the workplace impact on the environment. Within research facilities such as accelerator facilities, it should be considered as part of our role as members to incite this culture change.
Every single person has the responsibility and the agency to incite culture change. 

\begin{recommendation}\textbf{Enable and facilitate culture change from the top-down.}
Those with authority should create policies, provide access to knowledge transfer and training, budget the time and funding for efforts and \acrshort{randd} towards lower environmental impact, and lead by example in the workplace. Culture change can be progressed by educating and raising awareness, setting clear goals and objectives, involving stakeholders, creating incentives and rewards, and providing resources and support.
\end{recommendation}

\begin{recommendation}\textbf{Track, monitor and celebrate progress and achievements.}
\end{recommendation}

\begin{recommendation}\textbf{Encourage procurement of, or work towards, more sustainable designs.} This applies to those with any form of influence, such as those working with accelerator components and ancillaries.
\end{recommendation}

\begin{recommendation}\textbf{Encourage each and every person to take more sustainable day-to-day and long term actions} in the workplace and contribute to community engagement.
\end{recommendation}

%%%%%%%%%%%%%%%%%%%%%%%%%%%%%%%%%%%%%%%%
\subsection{Optimisation for sustainability alongside scientific output \label{sec:scientific_output}}

Large accelerator facilities have \textit{historically} been built with scientific output, not sustainability, as a priority, but the ecosystem can no longer support this methodology. A common question at the centre of sustainable research discussion is: ``At what point could or \textit{should} environmental sustainability outweigh scientific output?'' Answers to this question are dependent on many factors, including: 1) who one asks, 2) which research one discusses, 3) which timescale one considers, 4) whether financial costs are fixed or impacted positively or negatively, 5) what other impacts to the world and humanity it has, and 6) what the level of impact on sustainability compared to research is.

The climate crisis is a wicked problem, with complex social, political and environmental implications.
It can be argued that scientific research has wicked solutions\footnote{Here, a play by the authors on the wicked problem~\cite{wicked-problem}, in which scientific research has the potential to propose partial, evolving solutions to a wicked problem, which are multi-dimensional, interdependent and have value conflicts themselves.}, with complex social, political and environmental implications. 
Scientific research does not solely have wicked solutions to the climate crisis.
It has a multitude of solutions to many problems, wicked or not.
With this logic, it can be imagined that no scientific research would ever prioritise environmental sustainability over the scientific output it produces, as scientific research has other goals (note in the case of research on environmental sustainability this is a little paradoxical).
On the contrary, the answer to the question of whether environmental sustainability should ever outweigh scientific output should sometimes be ``yes''. 
This question should be asked regularly and considered in every scenario. 
When under consideration, include the three pillars of sustainability, supplemented with the concept of sufficiency\footnote{``Sufficiency policies are a set of measures and daily practices that avoid demand for energy, materials, land and water while delivering human well-being for all within planetary boundaries''~\cite{IPCC2022}}, to keep the discussion open and ever-evolving. 
An example case could be when considering some form of an accelerator upgrade. When the increase in impact on the environment is multitudes higher than the gains achieved in scientific output, the answer to the proposed question may be ``yes''.
When the time saved in operations to meet a scientific goal would reduce the environmental impact, even after full embodied carbon costs are taken into account, the answer may be ``no''.

Ultimately, nowadays, consideration of environmental impacts should be of high priority. 
Efforts should be taken to investigate whether a different, more sustainable, approach can achieve the necessary science output. In the field of accelerator science, this could potentially manifest itself as an alternative accelerator-based solution or a radically different scientific technique. Reference~\cite{SustainabilityConsiderations} discusses potential energy efficient accelerator concepts. 

\begin{recommendation}
\textbf{Consider where environmental sustainability may outweigh scientific output in your project}, following the above paragraphs.
\end{recommendation}

\begin{recommendation}
\textbf{Evaluate the possibility of performing the same science with a more compact accelerator} e.g. with higher gradient and larger magnetic fields. This could include either existing accelerator concepts with updated technologies, or newer accelerator concepts~\cite{SustainabilityConsiderations}. Reference~\cite{FutureColliderComparison} specifically refers to the potential of compactness of wakefield accelerators.
\end{recommendation}

In addition, efforts that are made to increase the productivity or efficiency of an accelerator facility should result in the facility consuming less resources, e.g. consuming less electricity or needing less run time to achieve the same science output. It should not result in the facility consuming the same resources to maximise the scientific output beyond the initial design and irrespective of balance. Here, to supplement sustainability, the concept of sufficiency could be examined.

It is also possible that the pursuit of maximal efficiency in an accelerator results in the `rebound effect'\footnote{The rebound effect (also termed the `Jevons paradox' as it was brought to light by William Stanely Jevons through his mathematical applications to economics, and examination of the coal industry) is the phenomenon where technological improvements that lead to an increase in efficiency result in an absolute increase in overall resource consumption, rather than reducing consumption (as one might expect)~\cite{Rebound_effect}. In the case of a particle accelerator one can envision this effect through an increase in beam luminosity design leading to a complete dataset being collected earlier than planned. Here perhaps a lower amount of resources would be consumed compared to the original design. However, rather than the accelerator baseline run time being reduced to the time taken to complete the dataset, instead more data was collected and therefore a higher amount of resources were consumed overall.}, whilst over-optimisation may induce challenges in adaptation to changing or unforeseen circumstances. Thus adaptability is a key component of consideration in the quest for increased environmental sustainability, particularly in resilience in the face of new or evolving situations.

\begin{recommendation}\textbf{Explore beam parameter optimisation} e.g. frequency regime, beam size, etc. Answering questions like ``Can an increase in luminosity optimise the science output per Watt of energy used?'' -- alongside the appropriate studies -- could lead to reduced resource consumption in the long-term. A reminder here to ensure the rebound effect is avoided. 
\end{recommendation}

\begin{recommendation}\textbf{Explore the potential for new/optimised technologies.} Everything from components to computation to materials; there is potential for technologies to aid in reducing overall environmental impact.
\end{recommendation}

\begin{recommendation}\textbf{Procure high efficiency components} e.g. klystrons, power supplies, variable speed drives, etc.~\cite{SustainabilityConsiderations}. 
\end{recommendation}

%%%%%%%%%%%%%%%%%%%%%%%%%%%%%%%%%%%%%%%%
\subsection{Prioritisation of impact reduction \label{sec:prioritisation}}

In general, the greatest ability to influence the whole-life environmental impacts of a large accelerator facility occurs at the conception, optioneering and design stages~\cite{BSIPAS2080}.
The most beneficial actions towards reducing the environmental impact would need to be designed into the facility itself. 
At these crucial stages, environmental impact assessments should be employed to highlight the areas of highest potential for environmental impact reductions and more sustainable designs (Section~\ref{sec:evaluation}). 

\begin{recommendation}\textbf{Embed sustainability into planning, design, construction and every other stage of the accelerator facility lifecycle.} 
At every design stage, from conception through to full technical design, decisions are made about the machine layout, beam energy, current, technologies and so on. Those decisions will take into account many different factors including scientific performance, reliability, technology readiness level (\acrshort{trl}), effects on other subsystems, cost, risk, and so on. Sustainability needs to be included in this list.
This is essential for enabling future operational and decommissioning environmental impact reductions. Thoughtful planning ensures responsible sourcing of resources (Section~\ref{sec:responsible_sourcing}) and optimal resource consumption reduction (Section~\ref{sec:resources_consumption}), laying the groundwork for a facility that prioritises sustainability across its lifecycle.

Here the community should standardise accounting and adhere to uniformity of reporting to ensure fairness and motivate the uptake of these practises~\cite{LDG-ESPPU}.

It will not be possible to always make the `best' decision in terms of environmental benefits, since there are so many other factors influencing design decisions, but the environmental impact of decisions made during the design process should be given an appropriate weight alongside other factors. To determine an appropriate weighting, there are multiple frameworks and analyses available including: the Multi-Criteria Decision Analysis (\acrshort{mcda}) in which weights are designed for conflicting criteria~\cite{MCDA}; Environmental, Social, Governance (\acrshort{esg}) in which factors are scored and aggregated~\cite{ESG}; Cost-Benefit Analyses (\acrshort{cba}) in which factors are converted into monetary terms~\cite{CBA}. As any form of weighting will have its own limitations, due diligence and transparency is absolutely necessary. \textit{Example case study: a social CBA of \acrshort{hllhc} in preparation for the FCC~\cite{FCC-CBA}, followed by a CBA of FCC~\cite{FCC_sustainability}.}
\end{recommendation}

\begin{recommendation}\textbf{Sustainability should be included as one of the key decision factors, alongside financial cost and scientific output.} Sustainability should not be presented as an optional extra; opting for a less sustainable design should be an active choice by the funders and require explicit justification and recognition of the trade-offs of providing less budget for sustainability. When presenting a design option to funders, the most environmentally sustainable version should be presented, even if it incurs higher financial implications. 
\end{recommendation}

\begin{recommendation}\textbf{Prioritise funding for \acrshort{randd} that demonstrates new technologies with the potential to significantly improve facility sustainability.}
Such investments are crucial for driving innovation in energy efficiency, material optimisation, and waste reduction, enabling facilities to meet ambitious sustainability goals while minimising environmental impacts. Funding bodies play a pivotal role in this process and must be encouraged to allocate resources for these efforts. By fostering technological advancements, these investments ensure that sustainability remains a central consideration in facility development and maintenance.
\end{recommendation}

\begin{recommendation}\label{rec:globalcollaboration}\textbf{Optimise global collaboration.}
On a global scale, the construction and operation of facilities with similar scientific operations could be optimised. Communication between facilities to coordinate their long-term operational timelines could reduce overall resource usage while maintaining world-leading scientific research. While this kind of coordination already occurs to some extent due to shared causes, funding, expertise, and personnel, explicit international communication and collaboration on scheduling could amplify efficiency and embed environmental considerations within the global scientific community. By viewing facilities as part of a cohesive network, rather than isolated entities, the global scientific community can achieve both environmental and operational excellence.
\end{recommendation}

\subsubsection{Weather events}
Many global and local problems will appear or be exacerbated as the climate crisis intensities. In relation to the scope of this document, the operation of an accelerator facility in an increasingly unstable environment will need consideration.
One such example is instability of the climate itself. 

Whilst everything must be done to prevent further exacerbation of the climate crisis, there is already evidence that weather events are increasingly more intense at a higher rate than records indicate~\cite{metoffice, UCAR}. Thus it is acknowledged that, whilst climate crisis mitigation efforts must be prioritised, adaptation efforts should be also considered within the lens of sustainable construction and operation. Adaptation can improve sustainability through damage prevention and resource consumption reduction.

\begin{recommendation}\textbf{Design the facility to adapt to the expected prevailing weather conditions} such as extreme weather events due to the climate crisis, weather effects of newly averaging higher/lower temperatures, rapidly changing temperatures, unpredictable temperature patterns and anomalous extreme temperatures during operation and throughout the lifetime of the facility.
\end{recommendation}

\begin{recommendation}\textbf{Plan for extreme weather events that halt operation of the facility} due to its impact on the safety of employees.
\end{recommendation}

\begin{recommendation}\textbf{Plan for droughts} that reduce allowed water allocation during operation.
\end{recommendation}

\begin{recommendation}\textbf{Consider the potential of flooding} that could reduce water quality/cleanliness during operation and directly impact and damage the accelerator facility. Consider mitigation options, e.g. flood water removal pumps and drainage options, sustainable paving (preventing run-off), a raised accelerator, etc.
\end{recommendation}
        
\begin{recommendation}\textbf{Plan for power availability reduction or cessation} due to extreme weather events.
\end{recommendation}

%%%%%%%%%%%%%%%%%%%%%%%%%%%%%%%%%%%%%%%%%%%%%%%%%%%%%%%%%%%%%%
\subsection{Consideration of all environmental impacts}

Often the phrase ``environmental sustainability'' is interpreted as ``\acrshort{ghg} impact'' or ``carbon impact'' however, other environmental impacts must not be ignored. For example, in the planning of land use, the impacts of the facility on biodiversity of local fauna and flora should be evaluated to identify problem areas, then efforts should be taken to mitigate or reduce those impacts. Waste -- hazardous or not -- is another example, where waste impacts are not limited to global warming impacts, and must be considered (See Section~\ref{sec:waste}).
Ultimately, there are many environmental aspects that should be taken into account when relevant, including but not limited to: biodiversity, waste, toxicity, noise pollution, air pollution, light pollution, water use, acidification and ionising radiation. There are a wide range of positive actions that members of facilities can take towards mitigation of environmental impacts when considering the broader aspects of sustainability.

\begin{recommendation}\textbf{Consider all appropriate environmental impact factors within each evaluation.} Assessments to measure environmental footprint such as environmental impact assessments or \acrshort{lca}s~\cite{iso14040,iso14044} for a large accelerator facility's construction, operation and end-of-life could be used to evaluate the wide range of environmental impact factors broader than just CO$_2$e emissions and identify areas and opportunities for improvement. %One could also consider following a Social enterprise Balanced Scorecard model~\cite{OpenUni}.
\end{recommendation}

%%%%%%%%%%%%%%%%%%%%%%%%%%%%%%%%%%%%%%%%
\subsection{Evaluation of environmental impacts and efforts \label{sec:evaluation}}

The pursuit of the reduction of environmental impacts of a large accelerator facility requires a large amount of data to inform and develop strategies. 
Thus environmental impact evaluations at each stage of the design of the facility should be performed. 
In the early stages, these assessments will not have anything like the detail needed for a full \acrshort{lca}; they should be used primarily to find areas of highest potential for impact reduction, to give approximate figures for competing options, and to highlight critical gaps in knowledge. \textit{Example case studies: Multiple proposed facilities are undertaking early studies as parts of more comprehensive LCAs to benchmark and develop strategies to reduce environmental impacts~\cite{FCC_carbon_optimisation, FCC_sustainability, ruedi, ARUPstudyCLICILC, C3, ISIS-II_LCA}.}
Versions of environmental impact assessments such as simplified LCAs could be used to perform hotspot analyses to identify key impact areas to focus on, and later used again to evaluate the effectiveness of impact reduction efforts. 
Section~\ref{sec:LCA_publications} lists more exemplary LCAs and other impact reports performed for research facilities.

%%%%%%%%%%%%%%%%%%%%%%%%%%%%%%%%%%%%%%%%
\subsubsection{Transparency \label{sec:transparency}}

Transparency and public reporting of emissions are essential for a facility's ethical environmental practices. They provide accountability, leadership, and further knowledge of areas that need improvement. Examples of such reports and strategies are those from \acrshort{cern}~\cite{CERNreport2023,CERNreport2025}, \acrshort{desy}~\cite{DESYreport}, \acrshort{esrf}~\cite{ESRFreport}, and \acrshort{stfc}~\cite{STFCreport}.

The environmental impact has the potential to be reported as a function of expected scientific output (e.g. CO$_2$e per amount of beam power, precision, luminosity, beam time, etc.), such as the CO$_2$e emissions per Higgs boson produced~\cite{C3, NatureCo2Higgs}. The reporting facility should be aware of the potential of greenwashing and sciencewashing, and should therefore report on absolute impacts alongside any other reported impacts.

\begin{definition}\textit{Greenwashing} is ``the intersection of two firm behaviours: poor environmental performance and positive communication about environmental performance''~\cite{greenwashing} and should be avoided at all costs. Due to the nature of the public funding of large accelerator facilities, transparency, accountability and honesty in environmental sustainability are essential. In addition, the full lifetime environmental, economical, social and scientific impact should be weighed against each other and all decisions should be well motivated and communicated transparently.
\end{definition}

\begin{definition}\textit{Sciencewashing} is defined as the ``deliberate action to simulate scientific practices or quality assurance to deceive others''~\cite{sciencewashing}. Though this does not often apply to large accelerator facilities, sciencewashing also encourages misconceptions of the potentials and limits of science and must be avoided.
\end{definition}

\begin{recommendation}\textbf{Communicate directly and openly about all impacts of a science facility.} This would include economic impacts (growth, high value jobs etc.), societal impacts (improved technologies, medical treatments etc.), and environmental impacts (carbon emissions, scientific progress in carbon capture, etc.). 
\end{recommendation}

\begin{recommendation}\label{rec:scientificoutput}\textbf{Report environmental impact reductions due to scientific output separately from a facility's negative environmental impacts.} This includes their potential for long term impact reductions.
\end{recommendation}

\begin{recommendation}
\textbf{LCA functional units should represent the accelerator beam delivered as a result of a facility design, expressed as a defined quantity of beam energy delivered under specified operating conditions, with the assumed operational lifetime explicitly stated.} In the `Sustainability Assessment of Future Accelerators’ document published by the Laboratory Directors Group Sustainability Working Group as a submission to the ESPPU (2025) it is recommended that -- for accelerators -- the functional unit should specify:
\begin{itemize}
    \item ``which accelerator systems are considered and how to treat pre–existing systems, e.g. only the main accelerator or injection systems and pre–accelerators'',
    \item ``which programme phases are included, e.g. only considering the first run phase, or subsequent upgrades (which may or may not be part of a baseline project proposal)'',
    \item ``which supporting infrastructure is considered, e.g. surface buildings, cryogenic plants, computing centres, power stations, water treatment plants, or roads''~\cite{LDG-ESPPU}.
\end{itemize}
The apparent detail required to be represented in the functional unit here can be expanded upon within the scoping stage of the LCA; the essence would be included in the functional unit.
\end{recommendation}

\begin{recommendation}
\textbf{Accelerator facility LCAs should use the cradle-to-grave system boundary.}
\end{recommendation}

\begin{recommendation}\textbf{Evaluate power consumption.} 
Collect (granular) data of power consumption of the various areas/sectors, and request data from service providers. 
Evaluate resource use e.g. for computing, utilise the Green Algorithms project~\cite{GreenAlgorithms}.
\end{recommendation}

\begin{recommendation}\textbf{Implement sub-metering} with sufficient granularity that future modifications at system level can be evaluated.
\end{recommendation}

\begin{recommendation}\textbf{Understand the predicted sources of energy available to the facility} at the time of construction and throughout the lifetime of the facility. Over time, varying energy sources may affect the environmental impact of the facility and, if negative, indicate the need for reconsideration. For example, the World Energy Outlook Report 2025 indicates a level of uncertainty for future energy scenarios~\cite{WorldEnergyOutlook2025}. \textit{Example case study: The Green \acrshort{ilc} project considered the phasing out of grid power in favour of sustainable power plants for a 200-300\,MW facility~\cite{GreenILC}}.
\end{recommendation}

\begin{recommendation}\textbf{Compare different design options with an environmental lens} e.g. detail relevant environmental considerations when comparing (e.g tonnage of concrete used, embedded \acrshort{co2}, estimated water usage rate, etc.).
\end{recommendation}

\begin{recommendation}\textbf{Develop precise modelling of anticipated beam loss points, radiation levels and worst case scenarios} to reduce required shielding thickness as a function of location within the machine. Consider enabling lower amounts of shielding by permitting higher radiation levels by removing (relatively fast) worker access to areas that could be managed with slower access, robotic access/repair, etc.
\end{recommendation}

\begin{recommendation}\textbf{Develop internal structures that inform end users of their environmental (and financial) costs\label{rec:show_costs}} for example of their simulations to educate and to prevent unnecessary and/or endless simulations.
\end{recommendation}

\begin{recommendation}\textbf{Implement documentation and tracking methods.} Documentation and publication of disused materials and components will increase likelihood of reuse. 
\end{recommendation}

\subsubsection{Waste\label{sec:waste}}

\begin{recommendation}\textbf{Reduce material waste in the machining stage.\label{rec:machiningwaste}} Consider optimising machining to reduce scrap and reduce amount of machining. 
\end{recommendation}

\begin{recommendation}\textbf{Efficiently and adaptively design test stands, prototypes and final components.}
\end{recommendation}

\begin{recommendation}\textbf{Minimise radioactive and toxic waste} through optimised design and ease of segregation of materials.
\end{recommendation}

\begin{recommendation}\textbf{Utilise waste products} where possible. E.g. implement thermal waste recovery within district heating. At a minimum, build district heating functionality into facility design for future implementation. 
\end{recommendation}

\subsubsection{Location}

\begin{recommendation}\textbf{Re-use existing support and scientific infrastructure where possible.}
\end{recommendation}

\begin{recommendation}\textbf{Choose site location considering lowering long-term environmental impact,} e.g. siting near abundant renewable energy sources, near a railway station to enable user access, or near a town or city with district heating that utilises waste heat. \textit{Example case study: The SESAME synchrotron light facility was the world's first large accelerator facility to be fully powered by renewable energy.}
\end{recommendation}

\begin{recommendation}\textbf{Investigate the benefits of various landscaping techniques and initiatives.} 
Implementation of various location dependent building colour or albedo consideration, green wall and roofing techniques, and material considerations can reduce the urban heat island effect, increase biodiversity, and reduce environmental and visual impacts.
Implementation of location dependent water redirection or management techniques can improve biodiversity and reduce increasingly occurring climate-related issues. For example, permeable pavement can lead to reduction of run-off and therefore reduction of flood risk and related issues. In addition, permeable pavements also help mitigate the urban heat island.
\end{recommendation}

\begin{recommendation}\textbf{Investigate biodiversity initiatives} such as implementing green walls, green roofs, green landscaping, beehives and housing\footnote{There are many types of endangered bees (i.e. not the honey-bee) that are not aided through supplying beehives.}, bug hotels, bird and bat houses in and around experimental halls and offices. \textit{Example case studies: OpenSkyLab at CERN held a volunteering day helping build insect hotels and cleaning fields from invasive plants. Many UKRI sites have installed bird houses, ponds, and insect hotels.} Some biodiversity initiatives can be interpreted as net-negative emissions, depending on the evaluation method. In this case, extra care must be taken in transparency of reporting.
\end{recommendation}

\begin{recommendation}\textbf{Communicate with local community and businesses,} e.g. to foster environmental collaboration, engage local knowledge and reduce impacts through local and sustainable businesses. \textit{Example case study: The Green \acrshort{ilc} project considers utilisation of local, sustainable building resources~\cite{GreenILC}.}
\end{recommendation}

%%%%%%%%%%%%%%%%%%%%%%%%%%%%%%%%%%%%%%%%
\subsection{Responsible procurement of resources \label{sec:responsible_sourcing}}

The procurement of resources for construction and operation of a facility should be streamlined, optimised and transparent.
Every material used (e.g. steel, helium, niobium, etc.) has an associated impact on the environment. Not only do they have environmental impacts from raw material extraction to physical manipulation, they can also have social impacts including unsafe mining conditions and unfair trade. This should not be ignored during a facility's design and procurement stages; (environmental) impact assessments could be used to highlight potential areas for more sustainable choices. 

Examination of the environmental impact of material procurement pathways can cover as much as the journey of raw materials from extraction to the use at a facility, or as little as just the machining of materials on-site. Here due diligence should be conducted to understand and document where a material is sourced, and members of the facility should strive towards procuring from the least environmentally impactful location. In addition, as a facility is reliant on the compliance of those providing materials, components and services, etc., clear communication with suppliers about the facility's environmental requirements throughout the procurement process is important.

Where reuse and recycling of materials and components from other facilities (such as magnets or shielding) is found to be a less environmentally impactful option, reuse and recycling should be implemented where possible. If they are not possible, the facility should use the least environmentally impactful material possible.

In addition, procurement of other resources such as electricity is highly influential in the total lifetime impact of a facility.

\begin{recommendation}\textbf{Integrate sustainability criteria into tender processes and purchasing decisions.} Investigate responsible, local and ethical procurement (of resources, components and parts). Every financial decision should have sustainability as a consideration. Communicate the need for sustainable procurement with suppliers. In particular, consider critical materials such as Rare Earth Elements (\acrshort{ree}) and high environmental impact materials. Be aware of and follow regulation on conflict materials\footnote{Materials from conflict zones that are sold to finance armed conflict and human rights abuses.} such as tin, tantalum, tungsten and gold (3TG). Reference~\cite{IFAST_Sustainable_Accelerators_Report} suggests that in an ideal scenario, their production paths should be certified. This can have the added benefit of making manufactures and vendors more incentivised to be clearer about the impact or embedded carbon of their products.
\end{recommendation}

\begin{recommendation}
    \textbf{Understand the status of relevant critical materials.} Critical material extraction can have large environmental impacts depending on the material and the location of where it is sourced. Defined critical materials vary by country based on geological endowments, with a lot of overlap. For accelerator facilities, some of the major critical materials are: cobalt, helium, niobium, silicon, tantalum, tungsten and some REEs, e.g. samarium and neodymium.
    Critical materials are highly dependent on geopolitics and distribution or concentration. Supply is often geographically concentrated, and diversification is needed to mitigate supply risks. Even in light of diversification, the demand growth may continue dependence on more volatile sources of materials. At the time of writing, Russia is subject to metal sanctions that could complicate some legacy routes, production and supply of many critical materials are dominated by China, and US tariffs are impacting trade routes and could cause interruption or delay in supply. Environmental impacts are affected by the method and location of sourcing. Long-term planning like reducing reliance on critical materials (or even stockpiling volatile critical materials to guarantee that the environmental impact does not increase) could aid in the reduction of impacts of a material used over the lifetime of an accelerator facility.
\end{recommendation}

\begin{recommendation}\textbf{Examine the potential for the adoption of local low-carbon energy generation} (photovoltaics (\acrshort{pv}), wind, hydro and nuclear small modular reactors (\acrshort{smr}) etc.). Besides relying on regional energy sources, facilities may be able to reduce their environmental impacts here, depending on embedded emissions. All opportunities in this space should be evaluated on a case by case basis with communication and consultation with local authorities. \textit{Example case studies: ``CERN signs long-term solar power agreements''~\cite{CERN-PV} and ``Utilization of renewable energies for sustainable accelerator operation at \acrshort{kit}''~\cite{gethmann:ipac2025-thpb096}.}
\end{recommendation}

\begin{recommendation}\textbf{Evaluate the potential of local power storage.} Consider the potential for the implementation of on-site batteries to store power from the electricity grid (and local sources) when it has a lower emission factor.  Redistribute for use at the facility when the electrical grid has a higher carbon intensity. Consider the full lifecycle of the batteries, especially in relation to raw material sourcing and later disposal.
\end{recommendation}

%%%%%%%%%%%%%%%%%%%%%%%%%%%%%%%%%%%%%%%%
\subsection{Resource consumption reduction \label{sec:resources_consumption}}

Accelerators are fundamentally resource consumptive. 
Some examples are: consumption of fossil fuels in the removal of tonnes of tunnelling material; using concrete as shielding which results in the emission of large amounts of \acrshort{co2e} from its production process~\cite{ARUPstudyCLICILC}; \acrshort{randd} towards new components often requiring multiple test stages and the machining of parts resulting in waste; the loss of increasingly scarce resources, such as helium~\cite{helium} through deteriorating systems~\cite{HeliumRecoveryMcGill}; and the consumption of electricity, particularly through cooling, cryogenics and the powering of magnets and radio-frequency systems. 
Currently, some of the world's largest accelerator facilities consume hundreds of \acrshort{mw} of wall-plug power, leading to annual energy consumption on the order of Terawatt-hours per year. 
For example, the European Spallation Source (\acrshort{ess}) has a maximum energy consumption estimation of $270\,$GWh/yr~\cite{ESSpower}, whilst the CERN laboratory consumes up to $1.3\,$TWh/yr depending on its operating regime~\cite{CERNpower}.

The generation of electricity, whether renewable, nuclear or fossil fuel generation, has an inherent environmental impact. No current form of electricity generation has a net zero or net-negative environmental impact, and extrapolations indicate that they will not meet net zero by 2050 without help from other forms of carbon offsetting\footnote{For example, the UK government Green Book toolkit for \acrshort{ghg} emissions for appraisal provides analysts electricity emission factors with an optimistic plateau at 0.002\,kg\ce{CO2}e/kWh by 2050 (data table 1 of Reference~\cite{UKgovGreenBookSupp}). Other estimates for the UK, for example calculated from Reference~\cite{UKgovElectricityEmissions} (``Annex A: Net Zero Strategy categories: Greenhouse gas emissions by source'' and Figure 4.1 data from ``EEP web figures: 2023-2050''), indicate factors upwards of 0.03\,kg\ce{CO2}e/kWh up to 2050.}.
In addition, the required shift to renewable energies and the current rate of uptake will result in an `energy gap', where supply will not be able to meet global growing energy demands~\cite{HECAPplus}, thus global energy demands must be reduced. Furthermore, when `green' energy is purchased, it often simply shifts the consumption of fossil fuel energy to another end user - the environmental impact of energy consumption therefore must be calculated from the total impact of a grid supplier. Purchase of green energy would influence the market and indicate demand for more green energy, but this is not enough. 
Future and current large accelerator facilities must therefore consume as little energy as possible.

Each facet of accelerator design has the potential for improvement in efficiency and reduction in resource consumption. This will only be achieved if efforts are taken in that direction. This is intrinsically linked to culture change (Section~\ref{sec:culture}), and prioritisation and planning (Section~\ref{sec:prioritisation}). To aid in the reduction of resource consumption, large accelerator facilities should perform environmental impact assessments to identify areas of potential reduction in the design stages.

\subsubsection{The accelerator facility}

\begin{recommendation}\textbf{Design the facility with waste reduction efforts implemented from the start} e.g. waste thermal energy used for district heating, gas recovery systems etc. \textit{Example case studies: \acrshort{ess} implemented waste heat recovery with Alfa Laval and with SSAB Zero TM steel heat exchangers~\cite{ESS-waste-heat,ESS_green_accelerator}. The LHC has implemented a waste heat recovery system validated with a digital twin~\cite{LHC-waste-heat, LHC-waste-heat-article}, estimated to save 25-30\,GWh/yr from 2027. Brookhaven National Lab (\acrshort{bnl}) presented potential for 43\,kt\acrshort{co2e} of avoided emissions through heat recovery and disuse of the BNL steam plant\cite{ESSRI_2024}.}
\end{recommendation}

\begin{recommendation}\textbf{Reduce gross material and resource use through efficient and optimised design and machining techniques.} Innovations in high- and ultra-precision machining could improve sustainability~\cite{machining-innovation}. See also Recommendation~\ref{rec:machiningwaste}.
\end{recommendation}

\begin{recommendation}\textbf{Implement resource recovery schemes} such as helium, helium 3, \acrshort{ghg}s, etc. \textit{Example case study: Helium Recovery at ISIS~\cite{HeliumRecoveryISIS}.} 
\end{recommendation}

\begin{recommendation}\textbf{Reduce water waste.} \textit{Example case studies: CERN has committed to keeping its water consumption up to the end of Run 3 within 5\% above their baseline year of 2018, (3,651 megalitres) despite the growing demand for water cooling at upgraded facilities through improving water infrastructure such as the North Area cooling tower recycling plant, the Antiproton Decelerator cooling infrastructure, and the demineralised water production units. 
Automated systems for monitoring and management were also implemented~\cite{CERNreport2025}}.
\end{recommendation}

\begin{recommendation}\textbf{Reduce power consumption.} Decarbonisation of national grids and local power supplies will only partially reduce the operational power impact of facilities, thus efforts towards higher efficiency and lower power consumption are still vital. Example case studies are provided in the following sub-sections.
\end{recommendation}

\begin{recommendation}\textbf{Increase efficiency of wall-plug to beam output.} Consider power efficiency at each step of power conversion continuously.
\end{recommendation}

\begin{recommendation}\textbf{Identify optimal accelerator operation regimes and implement demand shifting} (for each accelerator area) e.g. operational modes, $\beta$-function regimes etc. Consider greener power generation (e.g. seasonal and daily fluctuations, operation for optimal PV, boosting reliability with battery power, etc.). 
Include R\&D into new technologies that could optimise these regimes. 
\textit{Example case study: ``CLIC Energy Load and Cost Analysis''~\cite{Fraunhofer}}. 
\end{recommendation}

\begin{recommendation}\textbf{Plan longevity and reliability} e.g. to prevent leakage and loss for lifetime of operation, e.g. in cooling processes.
\end{recommendation}

\begin{recommendation}\textbf{Consider widespread utilisation of intelligent systems, artificial intelligence, machine learning and other methods} e.g. for real-time updates of loss and leakage to enable fast rectification, for real-time beam optimisation, for real-time power consumption optimisation, for real-time accelerator mode switching etc. It is explicitly noted that these methods can consume a high amount of resources and therefore increase environmental impacts where they were intended to reduce them. Extra care should be taken to evaluate their lifetime environmental impacts before they are implemented.
\end{recommendation}

\begin{recommendation}\textbf{Investigate the effects of localising vs. globalising/centralising systems} e.g. water cooling, cryogenic systems, thermal energy recovery.
\end{recommendation}

\subsubsection{Buildings}
\begin{recommendation}\textbf{Optimise building construction methods}. Utilise sustainability certified construction companies. Choose building materials optimal for their lifetime use case and plan for reuse, etc. Consider bio-materials for construction (e.g. bamboo~\cite{Bamboo_construction, Bamboo_bonding} has been shown to have potential in construction applications and could reduce environmental impacts).
Investigate and minimise the carbon impact of materials used through, e.g. 1) the amount of materials, 2) the type of materials, 3) the transport of materials, and 4) the material manufacture and construction methods.
\end{recommendation}

\begin{recommendation}\textbf{Consider implementing shorter-term construction phases} to reduce site disturbance, lower energy consumption, reduce material wastage and minimise noise and air pollution. Lower construction costs should be balanced against functional longevity, reliability and resilience.
\end{recommendation}

\begin{recommendation}\textbf{Reuse and recycle existing buildings} (where environmentally optimal) with priority on reuse, as larger carbon savings are generally made here~\cite{Building-reuse}. The whole lifecycle should be considered, e.g. the refitting of a building and operational impacts should be compared to the construction and operation of newer, more efficient buildings. 
\end{recommendation}

\begin{recommendation}\textbf{Reduce or replace rare metals and high Global Warming Impact (\acrshort{gwi}) materials} such as beryllium in relevant areas of the facility, e.g. RFQs, targets, etc.
\end{recommendation}

\begin{recommendation}\textbf{Remove the use of high GWI gases} such as F-gases. Where not possible, minimise their use.
\end{recommendation}

\begin{recommendation}\textbf{Remove consumption of environmentally impactful chemicals} e.g. \acrshort{pfas} and ‘forever chemicals’, plastics, etc.
\end{recommendation}

\begin{recommendation}\textbf{Maximise building efficiency} e.g. insulation, heating and cooling (\acrshort{hvac}), district heating, etc.
\end{recommendation}

\begin{recommendation}\textbf{Explore operational cooling options} such as less environmentally impactful operational cooling by:
\begin{itemize}
    \item optimising the absolute temperature of the accelerator tunnel, cooling water and user area.
    \item changing the accepted ranges of temperature stability. \textit{Example case study: The \acrshort{ruedi} facility found that a significant fraction of operational footprint is from management of heat loads. Further investigation on footprint reductions without performance is underway~\cite{ruedi}.}
    \item implementing other cooling methods, e.g. free cooling.
\end{itemize}
\end{recommendation}

\begin{recommendation}\textbf{Implement eco-building standards} (\acrshort{breeam}~\cite{BREEAM}, etc.) to optimise energy consumption and reduce environmental impacts. Specifically, build experimental halls with environmental consideration. \textit{Example case study: ESS constructed BREEAM `Outstanding' rated office buildings~\cite{ESS-BREEAM}}.
\end{recommendation}

\begin{recommendation}\textbf{Develop a waste handling programme} that encompasses waste reduction, waste recovery and waste redefinition to enable donation programmes, e.g. accelerator components, scrap materials, computer waste, etc.
\end{recommendation}

\begin{recommendation}\textbf{Investigate prevention and reduction of radioactive waste.}
\end{recommendation}

\subsubsection{Tunnelling}

\begin{recommendation}\textbf{Optimise the physical size of the accelerator} and tunnel, e.g. minimising the tunnel size or optimising the tunnel shape~\cite{Tunnel-design} could reduce overall environmental impact. The length of a machine is approximately proportional to the construction cost, including the tunnelling \acrshort{co2} emissions. In addition, cross-section of the machine and therefore tunnel also has an impact. \textit{Example case studies: Life cycle assessment of CLIC \& ILC. Opportunities for carbon reduction techniques were identified including reducing thickness of precast concrete segmental lining, and replacing shielding walls with concrete casing and earthworks fill obtained from tunnel excavation~\cite{ARUPstudyCLICILC}.} 
\end{recommendation}

\begin{recommendation}\textbf{Investigate tunnel excavation alternatives}  (cut-and-cover, etc.).
\end{recommendation}

\begin{recommendation}\label{rec:lowcarbon}\textbf{Utilise low(er) carbon construction materials} such as
\begin{itemize}
    \item alternative materials (e.g. to bulk materials like concrete and steel or high impact materials like some \acrshort{ree}).
    \item low-carbon concrete (either `green' concrete or alternate grades). \textit{Example case study: Arup published a potential 40\% embodied carbon reduction for CLIC and ILC through low-carbon concrete use~\cite{ARUPstudyCLICILC}}\footnote{Innovate UK recently invested £3.2 million into concrete decarbonisation~\cite{InnovateUKconcrete}.}. 
    \item responsibly sourced materials.
\end{itemize}
Consider the durability and lifetime within any overall lifetime impact comparisons, e.g. some materials may have lower CO$_2$e per kg than others, but need replacement or repair sooner. Other compromises may present themselves such as financial, which should be weighed accordingly (Section~\ref{sec:prioritisation}). 
\end{recommendation}

\subsubsection{Shielding}

\begin{recommendation}\textbf{Explore re-use of materials for shielding} from internal or external sources, e.g. re-use excavated tunnel material (potentially to be mixed within the planned concrete).
\end{recommendation}

\begin{recommendation}\textbf{Investigate substitution of materials within shielding}
    \begin{itemize}
        \item Gabion construction\footnote{A gabion is a volume filling cage or container of sorts, filled with various materials, e.g. rocks or sand.}.
        \item Recycled materials, e.g. iron shot.
        \item Non-toxic materials, e.g. tungsten, bismuth, iron compounds, and polymer-based composites.
    \end{itemize}
Consider the compromises that carbon reduction techniques may introduce (see Recommendation~\ref{rec:lowcarbon}).
\end{recommendation}

\begin{recommendation}\textbf{Investigate the use of other, more sustainable materials} (that may require more \acrshort{randd}) e.g. liquid-based shielding systems (e.g. water, molten salt), biological radiation absorption (e.g. for low radiation levels, fungal shields, plant-based shields, and biofilm layers), hybrid materials with nanotechnology (e.g. graphene-based materials and nano-structured lead alternatives), polymers with regenerative properties, and other radiation-resistant composites such as wood-cement composites~\cite{Wood_cement}. Consider the compromises that carbon reduction techniques may introduce (see Recommendation~\ref{rec:lowcarbon}). 
\end{recommendation}

\begin{recommendation}\textbf{Construct modular shielding} to optimise the potential for re-use or recycling in the future. 
\end{recommendation}

\subsubsection{General component design}

The authors would like to develop this section with the growing expertise and case studies in particular in relation to accelerator areas such as RF, magnets, etc.

\begin{recommendation}\textbf{Optimise/minimise the number of spare parts} through designing modular equipment with the same spare parts, through accurate simulation of use and lifetime and through designing for longevity.
\end{recommendation}

\begin{recommendation}\textbf{Improve reliability and reduce failure likelihoods.} Ensure to account for replacement and repair of components.
\end{recommendation}

\begin{recommendation}\textbf{Design so as to ease repair, reuse, refurbishment and disposal.} Implement modularity of components or materials for later re-use or recycling. Material separation i.e. materials that can easily be separated at end of life will enable, simplify and encourage more environmentally friendly processing~\cite{IFAST_Sustainable_Accelerators_Report}. Consider coating types and methods, and removal potential.

In particular, plan for deconstruction and the storage of radioactively hot components with tracking and processing of materials that are safe to recycle in shorter timescales. This should be performed with evaluation and discussion with the relevant legislative authorities considering the uniqueness and/or specificities of the facility.
\end{recommendation}

\subsubsection{Radio Frequency (RF)}
\acrshort{rf} technology is required to accelerate beams to high energies. There are many types of RF technology used within particle accelerators, including (but certainly not limited to) technologies for: RF acceleration such as RF cavities; RF bunching such as RF Quadrupoles; RF sources such as modulators and klystrons; RF loss prevention such as Fast Reactive Tuners (\acrshort{frt}).
For consideration of efficient RF sources, the reader is directed to the \acrshort{ifast} Sustainable Accelerators Report~\cite{IFAST_Sustainable_Accelerators_Report}. It is recommended that new accelerator projects in particular should focus on using efficient RF sources. \textit{Example case studies: \acrshort{ifast} produced a working prototype of a klystron with a beam-to-RF power of 70\% efficiency ~\cite{IFAST_Klystron}. \acrshort{ess} improved financial and environmental costs through efforts on stacked-multi-level modulators, multi-beam \acrshort{iot}s, and klystrons~\cite{ESS_green_accelerator}. The HL-LHC implemented an upgrade to high efficiency TH2176 klystrons~\cite{TH2167_Klystron}. However, it is noted that these higher efficiency klystrons require about the same power as those pre-upgrade at the LHC, exhibiting an example of the Jevons Paradox.}

\begin{recommendation}\textbf{Procure or design RF technology to reduce impacts compared to current technology} e.g. thin film cavities, topology optimisation, solid state amplifiers, tuners, etc. \textit{Example case studies: Recent RF developments include the iSAS and SEALab team demonstrating the viability of a superconducting RF (\acrshort{srf}) photocathode~\cite{SRF-photoinjector}, and multiple teams working on thin-film SRF cavities~\cite{CERN-thinfilms, ASTeC-thinfilms}.} 

FRTs are a newer technology that reduce losses in RF cavities by enabling controlled changes in the cavity frequency through electronic control. Low-beam current Superconducting (\acrshort{sc}) machines could benefit from FRTs, including low-Beta accelerators or high current ERLs~\cite{FRTs}. \textit{Example case study: The Ferroelectric FRT (FE-FRT) is expected to be able to reduce peak forward RF power by about an order of magnitude, reducing
the wall-plug power consumption~\cite{FE-FRT}.}
\end{recommendation}

\subsubsection{Magnets}

The environmental impact of magnets is a widely discussed topic, with nuances in magnet type (e.g. permanent magnet versus electromagnet~\cite{PMs_Nature}, and warm (Normal
Conducting (\acrshort{nc})) versus SC versus High Temperature Superconducting (\acrshort{hts}) magnets~\cite{HTS_SC, IFAST_Sustainable_Accelerators_Report}). This version of this living document will not aim to provide definitive recommendations on which type of magnet is the least environmentally impactful, as currently there are many different use cases. Instead, conclusions of existing studies will be shared. For SC magnets and HTS magnets, the \acrshort{ifast} summary report from 2025~\cite{IFAST_Sustainable_Accelerators_Report} provides a comprehensive overview. 

\begin{recommendation}\textbf{Investigate different magnet technologies for each type of application} (e.g. permanent magnets versus electromagnets). \textit{Example case studies: The capital expenditure (\acrshort{capex}) and \acrshort{co2e} emissions of PMs is, in general, significantly higher than EMs, however the operational expenditure (\acrshort{opex}) and CO$_2$e emissions are negligible due to requiring no power (unless tuning is required) or cooling. For PMs -- that have a lifetime of 10 years or more -- the financial break-even point is under 5 years, and the environmental break-even point is after just 1 year of operation~\cite{PM_IPAC2023}. Variable Permanent Hybrid Magnets were proposed for BESSY III as an alternative to a completely fixed lattice or to requiring EMs, with power consumption of the storage ring having the potential to be reduced by over 0.5\,MW~\cite{PM-BESSY-III}. The \acrshort{ifast} project designed the Hybrid Electromagnet-Permanent magnet Tuneable Optics (\acrshort{hepto}) prototype for Diamond-II~\cite{HEPTO}. An Energy SAving Beam Line Magnets (\acrshort{esablim}) study showed that MgB$_2$ superconducting coils are effective for efficient beam line superferric cryogen-free magnets and could save energy by up to a factor of 40 compared to NC designs~\cite{ESABLIM}.} There are other environmental impacts to be considered here, which may outweigh the CO$_2$e savings, as discussed at the I.FAST workshop `Critical Materials and Life Cycle Management: The Example of Rare Earths - curse or blessing?' in 2023 (\url{https://indico.desy.de/event/35655/}). Here responsible sourcing\footnote{whilst not ignoring the intersection with land-connected peoples~\cite{land-connected-peoples}}, clean extraction, certified supply chains and methods of recycling should all be considered. 
\end{recommendation}

\subsubsection{Computing \label{sec:computing_construction}}

Data handling and processing, and the hardware required to provide these services, is a highly consumptive area that is necessary for accelerator facilities to function. In addition, there is a complex relationship between the resources required for hardware and operation that make the evaluation and reduction of environmental impacts complex themselves. In relation to the rebound effect, efficiency savings at the data-centre may not lead to reduced consumption as the demand for computing workload demands is often not reduced in tandem,
which can lead to fewer resources being used for a longer time to complete
work pledges. In addition, as electricity generation tends towards decarbonisation, the embedded carbon impacts of hardware become a larger fraction of overall computing carbon impact.

Due to economies of scale, it is usually more cost effective (both in terms of carbon and price) to aggregate the necessary resources in a central location close to the accelerator complex, and thus requiring an on-site data centre. It is possible that computing would be considered out of scope if the facility uses off-site services, depending on the environmental impact study, but if so it should still be acknowledged (and evaluated in comparison to the consumption of the facility). Outwith running more efficient work at data-centres, there are a number of things data centres can do themselves to reduce their impact. 

For example, the Worldwide LHC Computing grid's (\acrshort{wlcg}) sustainability strategy is to ``reduce the cost, energy consumption, and carbon footprint of computing by being able to use a wide range of resources including Grid, public or private clouds, HPCs, and by expanding the architecture.''\footnote{See \url{https://agenda.infn.it/event/44943/contributions/266400/attachments/137342/206352/Software_ESPPU_Skidmore.pdf}}.
They are working to agree on metrics and provide a framework to collect information related to energy efficiency. They will facilitate the use of more energy-efficient hardware where possible, and they will develop and promote a sustainability plan.

There are many comprehensive sustainable computing references that the author may find useful (e.g. Reference ~\cite{BPcomputing10}). This document aims to share recommendations directly relevant to accelerator facilities. 
The computing recommendations will be split into three sections, the first denoting best practises in purchasing and operating hardware, the second for best practises for running or designing software (to be run at computing centres and beyond), and the third suggesting the utilisation of compute power to improve sustainability.

\paragraph{Sustainable hardware practises\label{rec:sustainable_hardware_practises}}
\begin{recommendation}\textbf{`Breathable' computing centres.}
As computer centres have a power consumption that can be sculpted, it is possible that computing centres are able to scale power usage on demand somewhat. Related to demand shifting energy consumption, `supply shift' provision of computing resources by clocking down/up computing nodes or in rare cases powering off/on nodes. Data centres that are situated on grids with less clean energy portfolios have the potential to mitigate their carbon footprint by preferentially utilising greener energy sources for computational tasks; however, it should be noted that this measure does not intrinsically decrease total electricity usage.
\end{recommendation}

\begin{recommendation}\textbf{Evaluate and optimise hardware} where possible.
\textit{Example case study: Through operation impact evaluation, \acrshort{gpu} implementation at the \acrshort{cbm} experiment was shown to improve energy efficiency compared to \acrshort{cpu}s and therefore associated \acrshort{co2e} emissions during data processing~\cite{Kalman_Fitter_CO2}.}
\end{recommendation}

\begin{recommendation}\textbf{Right-size servers and devices.}
Order what you need when you need it. Talk to large user-groups to establish their needs in advance and order equipment to those specifications, to avoid hardware not being utilised for long periods of time. Notably, this could increase CapEx costs which would need to be taken into account in decision making.
\end{recommendation}

\begin{recommendation}\textbf{Make resources more widely available.}
\textbf{Practise open science with data preservation and data sharing~\cite{data_preservation}.} 
Enable and encourage the use of Fair Share Scheduling (\acrshort{fss}) where resource allocation is done by user group.

Ask if it is possible for external groups to make use of resources (allocated or otherwise) that are under-utilised? 

Allow work from places with electrical grids with higher emissions factors to run on spare local resources. \textit{Example case study: the Sustainable Federated Compute Infrastructures (\acrshort{susfecit}) Research Network project, planned in Germany seeks to adapt the amount of calculations and energy consumption to renewable energy supply\footnote{See: \url{https://indico.desy.de/event/48572/contributions/193591/attachments/99720/138103/susfecit-atlas-f2f-6oct2025.pdf}}}.
\end{recommendation}

\begin{recommendation}\textbf{Advertise hardware capabilities.}
Communicate to users what machines are available, when and their capabilities to allow users to optimise their code for specific machines.
Utilise online portals. Develop and facilitate training workshops.
\end{recommendation}

\begin{recommendation}\textbf{Standardise software and hardware} for the full lifetime of the facility.
Mitigate likelihood of (planned) obsolescence or vendor lock-in.

Perform a cost-benefit analysis of the use of similar machines. Replacement parts, like memory (which is the second most costly in terms of embedded carbon after the \acrshort{cpu}) could be cheaper to obtain and possible to be recycled from older machines. However, compatibility issues and overall efficiency could outweigh these benefits, as many modern compute elements have fixed (soldered or in package) memories, which are faster and more power efficient.
\end{recommendation}

\begin{recommendation}\textbf{Develop sustainability requirements for equipment procurement.}
For commercial tender processes, add requirements (and budget allowances) that help favour companies that provide evaluations such as life cycle assessments, e.g. information about the embedded carbon and other manufacture resource use (land, water, etc.).
\end{recommendation}

\begin{recommendation}\textbf{Consider utilising highly efficient (super)computers.} When evaluating providers of (super)computers, consideration should be made towards their power efficiency (e.g. Co-efficient of Performance (\acrshort{cop})) and their carbon impacts. More environmentally friendly providers should be prioritised, where possible.
\end{recommendation}

\begin{recommendation}\textbf{Apply for sustainability accreditation and utilise existing schemes} such as digital sustainability certification, e.g. Green DiSC~\cite{GreenDiSC}.
Use the accreditation as a public exercise to obtain the political leverage to inspire and create longer-lasting changes to computing operations.
\end{recommendation}

\begin{recommendation}\textbf{Develop or utilise existing `research obsolete computing hardware' donation programmes.}
\end{recommendation}

\begin{recommendation}\textbf{Write, and regularly review a data management plan.} There are significant environmental impacts from data storage and transfer. \textit{Example case study: The \acrshort{grand} project propose to drastically reduce the volume of data to be archived, and store data in countries with low energy emission factors~\cite{GRANDcarbonfootprint}.} 

In addition, the GRAND project showed an excellent example of how complete re-evaluation of the way we currently do things could lead to unexpected options to reduce environmental impacts. For example, it could even be the case that physically hard-copying and flying/shipping data around the world to the final cloud storage data centres could be less impactful than transferring data online, depending on electricity emission factors. This would naturally have other impacts that would need to be taken into consideration, such as direct plane emissions into the atmosphere, and labour costs.
\end{recommendation}

\begin{recommendation}\textbf{Optimise cooling infrastructure and techniques.} 
Computing cooling power and resource consumption is often a large proportion of compute impacts.
The overall energy efficiency of data-centres is commonly characterised by the Power Usage Efficiency (\acrshort{pue}), which is the ratio of the total energy used by the facility over the energy used by the active \acrshort{it} elements (compute, network, storage). Ideally this value would be unity. 
In practice modern data-centres can achieve values around or even better than $1.1$. Limitations come from the geographical location. The most important overhead consumption is the cooling of the data-centres, i.e. the extraction of the waste-heat (power-losses in transformers and cables are a factor but much smaller). 
The choice of the cooling technique is therefore essential. 
It must be kept in mind that some of the techniques have other ecological impacts usually from fresh-water or various coolants. 
\textit{Example case study: The \acrshort{alice} O$^2$ Event Processing Nodes (EPN) farm infrastructure uses adiabatic cooling system that is not only more energy efficient and environmentally friendly, but results in reduction of operational and capital financial costs~\cite{ALICE_HPC_eff}.}

In the best data-centre designs the waste-heat can be reused for something else (typically heating). Again different cooling techniques will result in the waste-heat being extracted into media (air, water) at different temperatures, the higher the temperature the more efficient and easy the re-use. Energy reuse is often very expensive or nearly impossible to retrofit and best included by design. 
\end{recommendation}

\paragraph{Sustainable software practises\label{rec:sustainable_software_operations}}

\begin{recommendation}\textbf{Use or create software for general use.}
When creating software, consider the potential environmental benefits (and impacts) of open-source implementation and follow the Findable, Accessible, Interoperable and Reusable (\acrshort{fair}) principles. The Data Preservation in High Energy Physics (\acrshort{dphep}) collaboration recommends that, alongside making HEP analyses FAIR, HEP should focus on preservation, which means keeping data alive, accessible and usable, across ever changing IT infrastructure and technologies~\cite{DPHEP}. \textit{Example case study: The SoftWare InFrasTructure for High Energy Physics (\acrshort{swifthep}) project~\cite{SWIFT-HEP} (with SWIFT-HEP2 proposed to continue efforts in the future) is a key contender in the international roadmap for the development of software for HEP experiments, including efforts towards efficient and sustainable event generation for the \acrshort{hllhc}~\cite{Bothmann_2022}}.
The Software Sustainability Institute provides a wealth of resources relating to all pillars of sustainability~\cite{SoftwareSustainabilityInst}.
\end{recommendation}

\begin{recommendation}\textbf{Consider the programming language used.}
Reference~\cite{astro-computing-impact} suggests reconsidering standard Python (commonly used within Physics) towards either the use of high-performance libraries or transitioning towards a more environmentally friendly (compiled) alternative.
\end{recommendation}

\begin{recommendation}\textbf{Optimise Continuous Integration and Continuous Delivery (\acrshort{ci/cd}) pipelines.} Reference~\cite{CDCI-github-impact} highlights the environmental impacts of implementation of CI/CD pipelines, such as GitHub Actions, and provides calls to action which we draw from here. 
Facilities and users should evaluate CI/CD providers, the regions where their runners are deployed, and their resource consumption reduction actions~\cite{Optimization-github-impact}. 
The provider and developers should examine their use of scheduled runs (particularly on GitHub forks), and consider disabling scheduled runs by default and after an optimised number of days of inactivity~\cite{CDCI-github-impact}. 
Developers should also reduce repository sizes where possible and minimise unnecessary files~\cite{CDCI-github-impact}.
\end{recommendation}

\begin{recommendation}\textbf{Inform end users of their impacts.} See Recommendation~\ref{rec:show_costs}.
\end{recommendation}

\begin{recommendation}\textbf{Provide a common simulation data library} to limit the amount of simulations users run. \textit{Example case study: The GRAND project~\cite{GRANDcarbonfootprint}}\footnote{Other institutions have implemented similar schemes, however the GRAND Project explicitly mention this as a solution to reducing environmental impact.}.
\end{recommendation}

\begin{recommendation}\textbf{Invest in more efficient software} either through rewriting/restructuring existing codes, or financially investing in more efficient software.
\end{recommendation}

\begin{recommendation}\textbf{Provide access to the right tools to do analyses.} \textit{Example case study: STFC Scientific Computing produced `FitBenchmarking’, an open source Python package that compares data fitting software to enable researchers to save both time and energy thorough identifying the optimal ``fitting'' algorithm for their analyses~\cite{FitBenchmarking}.}
\end{recommendation}

\paragraph{Computing for sustainable practises}
\begin{recommendation}\textbf{Develop a virtual accelerator (digital twin)\label{rec:digitaltwins}} for tests of particle accelerators and components with a focus on optimising for sustainability. Include the ability to evaluate environmental impacts within digital twins. 
\textit{Example case study: \acrshort{rf2.0} project on AI-assisted power management focusing on ``optimised operations during beamtime, accelerated start-up and shut-down operations, digital twins for accelerator design and operations''~\cite{RF2.0}.}

In addition, digital twins could aid studies further than energy consumption, such as allowing consideration of beam optimisation with an environmental lens.

Consider the benefits and environmental impact of digital twins themselves - simulation often comes with large compute requirements.
\end{recommendation}

\subsubsection{Travel}
\begin{recommendation}\textbf{Decarbonise and improve options for local daily travel methods of staff} e.g. understand the needs and consider provision of local or subsidised transit. Communicate these needs and collaborate with the local government travel sector.
\end{recommendation}

\begin{recommendation}\textbf{Reduce air travel.} At 96\% of the GHG emissions of research travel, air travel is the environmental impact with most potential for reduction~\cite{Flight_quotas}. Only use air-travel when necessary e.g. to provide inclusive opportunities to all staff. Consider implementing an air-travel miles budget. \textit{Example study: ``Flight quotas outperform focused mitigation strategies in reducing the carbon footprint of academic travel''~\cite{Flight_quotas}.}
\end{recommendation}

\begin{recommendation}\textbf{Implement remote/hybrid user access} in relevant areas of the facility, e.g. to minimise the need for large groups of users to visit the facility to carry out an experiment and therefore encourage reduction in travel, and to provide additional (international) person-power for different types of facility shifts or projects over all time zones.
\end{recommendation}

\begin{recommendation}\textbf{Optimise meetings and conferences.} 
Enable and encourage remote/hybrid meetings/conferences. Consider the implementation of fully online versions of large conferences. Consider implementing local conference hubs for large international conferences. Reduce in person meeting/conference regularity\footnote{These suggestions do not need to be implemented for every edition of a conference, but could be implemented into rotation.}. Remove the expectation and necessity for scientists to travel to (overseas) meetings/conferences. 

Remote meetings and asynchronous work settings perform well for status updates, routine reporting, documentation, execution planning, meetings where one speaker is mostly presenting, data review (that uses screen sharing), and information transfer. In-person meetings are beneficial (but not necessary) for clarity (when the cost of misunderstandings can be high), creativity, networking, and collaboration initiation. 
\textit{Example study: Virtual conferencing can reduce carbon footprint by up to 94\% and energy use by 90\%~\cite{Virtual_and_Hybrid}}.

As learnt from the COVID-19 pandemic, online conferences and meetings are feasible at the large scale. However, from 2022 to 2024 there has been a 70\% reversion to in-person conferences~\cite{Reversion_To_In_Person}. \textit{Example study: online/hybrid conference strategies reduce carbon footprint and increase accessibility~\cite{Reversion_To_In_Person}.} 
\end{recommendation}

\begin{recommendation}\textbf{Encourage lengthening/optimising in-person stays~\cite{GRANDcarbonfootprint}.}
\end{recommendation}

%%%%%%%%%%%%%%%%%%%%%%%%%%%%%%%%%%%%%%%%%%%%%%%%%%%%%%%%%%%%%%%%%%%%%
\subsection{Carbon offsets \label{sec:offsets}}

Carbon offsets should only be used or purchased as a last resort when other alternatives to prevent and reduce carbon impacts have been exhausted. The legitimacy of many carbon offsets is under critique~\cite{carbonoffsets}, so carbon offset regulation should be considered effective before any implementation~\cite{netzero}. 
In addition, risk of reversal\footnote{Here, reversal is where sequestered carbon is re-released back into the atmosphere.} and durability level\footnote{Durability of carbon offsetting methods is how long the carbon stays locked away before being re-released back into the atmosphere.} of carbon offsetting methods can have significant impacts of its effectiveness~\cite{OxfordCarbonOffsetting}. 
However, this does not mean that measures should not be taken to aid in carbon sequestration, which is a necessary action to reduce existing \acrshort{co2} in the atmosphere.

\begin{recommendation}\textbf{Implement net-negative measures.} 
Net-negative measures ultimately mean more CO$_2$e is absorbed than emitted which can be considered as carbon offsetting. Implement measures to reduce the risk of reversal and increase durability.
Plant trees (in optimal positions and of optimal types) on-site.
Utilise materials that sequester CO$_2$ and other GHGs.
Take care with transparency in reporting net-negative emissions. 
\end{recommendation}

%%%%%%%%%%%%%%%%%%%%%%%%%%%%%%%%%%%%%%%%
\subsubsection{The environmental impact of science output \label{sec:scienceimpact}}

The justification of scientific output as a reason for high energy consumption, carbon emissions and extractive processes, can be interpreted as `greenwashing', and is thus a delicate topic. 
In addition, the societal benefits versus impacts are often ill-defined and unclear to the general population. 
Facility environmental impacts cannot be directly cancelled out by their scientific progress/outputs. This is because once \acrshort{co2e} is emitted to the environment, it stays there for hundreds to thousands or years, or until it is captured. Even if, as a direct result of an accelerator facility's research, highly efficient Carbon Capture and Storage (\acrshort{ccs}) technology is developed that would mitigate the CO$_2$e emissions produced in the facility's construction, operation and decommissioning, that CO$_2$e has already been emitted to the environment and must be reported and accounted for transparently.
See Recommendation~\ref{rec:scientificoutput}.

Though the impacts of a facility cannot be offset by scientific output, the scientific research performed there can still contribute significantly towards research in sustainability. 

\begin{recommendation}
\textbf{Encourage scientific research that works towards sustainability.} This role is considered crucial. Accelerator facilities should pursue and promote knowledge exchange to both mitigate and adapt to the climate crisis. To encourage this, consider awarding recognition to e.g. users at a user-based facility whose research contributes towards the three pillars of sustainability. Consider the impacts of implementing a quota of research in sustainability projects that the facility will achieve in a given user run.
\end{recommendation}

%%%%%%%%%%%%%%%%%%%%%%%%%%%%%%%%%%%%%%%%
\subsection{Skill sets, knowledge transfer, and interdisciplinary collaboration\label{sec:skillsets}}

Various sustainability skill-sets are required for the different phases of construction, operation and decommissioning of a large accelerator facility; it is important to define these skill-sets and ensure they are embedded in the facility's resources, through the upskilling of members of staff and/or through the hiring of experts. The collective sharing of knowledge through the presentation and publication of efforts and results -- if only to teach others of mistakes to avoid -- is essential. 
%Save doubling up on efforts/reinventing the wheel 

\begin{recommendation}\textbf{Develop education programmes and resources for relevant stakeholders} e.g. sustainability best practices for staff, environmental impact estimation for staff, sustainability efforts of the facility for the general public etc.
\textit{Example case studies: sustainable computing workshops within the High Energy Physics department at DESY~\cite{Sustainable_Computing_DESY}, and the Erforschung von Universum and Materie (\acrshort{erum}) (Research on Universe and Matter) Data Group who have run a space for ErUM-Data communities, to interact and share information with one another~\cite{ErUMDataGroup,ERUM_sustainability_ethics} and created an action group on Sustainability and Ethics in ErUM-Data}. See Section~\ref{sec:courses}.
\end{recommendation}

\begin{recommendation}\textbf{Provide adequate environmental impact evaluation software and databases to relevant stakeholders.}
\end{recommendation}

\begin{recommendation}\textbf{Perform evaluations of programmes, resources, etc.}
The community could develop governance and define evaluative frameworks to determine the efficacy of implemented training, tools, resources, etc. This should be done on an iterative basis, with surveys and key performance indicators being taken as intermediary measures. Structures could be developed to aid the community in the evaluation. \textit{Example case studies: The organisers of the inaugural sustainable computing workshops at the HEP department at DESY provides details on their structure, success and plans for the future~\cite{Sustainable_Computing_DESY}. Within astronomy, the Fast Radio Burst 2025 conference shared their experience of building an inclusive and accessible hybrid conference, including their sustainability efforts and results~\cite{FRB_conference}. The LDG Sustainability Working Group developed guidelines and a list of key parameters for the assessment of the sustainability of future accelerators in particle physics, whilst also providing a summary of sustainability best practices~\cite{LDG-ESPPU}.}
\end{recommendation}

\begin{recommendation}\textbf{Communicate with policy-makers} to educate on the requirements of a facility that is trying to reduce its environmental impact, e.g. the requirement of disposal rather than storage of radioactive materials could limit likelihood of reuse or recycling.
\end{recommendation}

\begin{recommendation}\textbf{Build interdisciplinary collaborations} to drive knowledge transfer and scientific cooperation~\cite{Mega_RIs}. \textit{Example case studies: some medical linac LCAs have shown that idle or `standby' power is a large proportion of their carbon footprint (8\%-19\%), second only to patient travel~\cite{Medical_carbon_footprint_UK, Medical_carbon_footprint_US}. Whereas in another study, accelerator acquisition and maintenance was the major carbon impact at 37.8\%~\cite{Medical_carbon_footprint_Fr}. This exhibits the variation of impacts and dependencies between scope and location. Communication and collaboration with ~\cite{Mega_RIs}.
In addition, medical linacs suffer similar sustainability issues compared to physics linacs, such as leakage of SF6 (with an average leakage rate of 0.14\,kg\,CO$_2$e/yr/linac) ~\cite{Medical_carbon_footprint_UK}. Common issues could benefit from common solutions found through collaboration.
In addition, efforts are being made to reduce energy consumption of medical linacs~\cite{Software_Linac_Energy}, which again could open opportunities for collaboration.
}

Include collaborations with humanities and social sciences to help answer the complex interdisciplinary questions raised in the quest for sustainability. This includes building understanding of local contexts, histories, and ultimately going beyond the traditional expertise of physicists, accelerator researchers, etc.~\cite{GreenPhysics_SocialSciences}. \textit{Example case study: Maunakea is the proposed location of the Thirty Meter Telescope. However, conflicts arose from a disconnect between the Native Hawaiians, Hawaii residents, and the international astronomy community, discussed by Native Hawaiian natural scientists and allies in Reference~\cite{Maunakea}. This includes discussion of the relation between Native Hawaiians and the concept `aloha ‘āina' -- a familial love for and commitment to sustaining the land~\cite{Maunakea}.
Interdisciplinary expertise is needed to enable listening,  learning, and dialogue between communities to build long-term and sustainable partnerships inherently based on respect and trust~\cite{IndigenousEngagementInClimate}.}
\end{recommendation}

\begin{recommendation}\textbf{Collaborate with industry, stakeholders, partners and suppliers} to promote knowledge transfer and drive new research and innovation. See also Recommendation~\ref{rec:globalcollaboration}.
\end{recommendation}

%%%%%%%%%%%%%%%%%%%%%%%%%%%%%%%%%%%%%%%%%%%%%%%%%%%%%%%
\clearpage
{\color{gray}\hrule}
\section{Additional Resources \label{sec:resources}}
{\color{gray}\hrule}
\vspace{1.5em}

This section collates resources that may be beneficial towards evaluating and reducing the environmental impact of large accelerator facilities.

\subsection{Organisations}
\begin{itemize}
	\item For examples of institutional, group and individual actions that could be adopted, the Sustainable \acrshort{hecap+} reflection document ``Environmental sustainability in basic research: a perspective from HECAP+'' published in 2023~\cite{HECAPplus} provides recommendations in areas of computing, energy, mobility, research infrastructure and technology, food, resources and waste.

    \item The \acrshort{icfa} Panel on Sustainable Accelerators and Colliders assesses and promotes ``developments on energy efficient and sustainable accelerator concepts, technologies, and strategies for operation, and assess and promote the use of accelerators for the development of Carbon-neutral energy sources.'': \url{https://icfa.hep.net/icfa-panel-on-sustainable-accelerators-and-colliders/} 

    \item Innovate for Sustainable Accelerating Systems (\acrshort{isas}) (2024-28)~\cite{iSAS} Horizon Europe project working on ``establishing enhanced collaboration in the field to broaden, expedite and amplify the development and impact of novel energy-saving technologies to accelerate particles'': \url{https://isas.ijclab.in2p3.fr/}.

    \item Flexibility in RIs for global CArbon Neutrality (\acrshort{flexrican}) (2024-27): three \acrshort{esfri} infrastructures, three academic institutions and two industrial partners from Europe working to demonstrate ``how research infrastructures as electro-intensive actors can enhance energy flexibility for the European electrical grid and contribute to local heating networks through Waste Heat Recovery projects.'': \url{https://flexrican.eu/} and 
    \url{https://cordis.europa.eu/project/id/101131516}.

	\item The Europe-America-Japan Accelerator Development Exchange Programme (\acrshort{eajade}) (2023-27)~\cite{EAJADE} has a work package for sustainable technologies for scientific facilities including high efficiency SC cavities, RF amplifiers and more~\cite{EAJADEkickoff}.
    
    \item Research Facility 2.0 (\acrshort{rf2.0}) (2024-26)\cite{RF2.0}: Towards a more energy-efficient and sustainable path. This Horizon Europe Research and Innovation project is designing a variety of products to be used at large research facilities to reduce their environmental impact. Tunable permanent magnets and novel solid state amplifiers for accelerators, \acrshort{pmu}s for real-time energy monitoring of local electrical grids, and digital twins for breathable data centres. The project aims to build 4 demonstrator projects to showcase these products in production environments. In addition it will also be investigating ways accelerator complexes can transition to greener power: \url{https://rf20.eu/} and 
    \url{https://zenodo.org/communities/rf20/records?q=&l=list&p=1&s=10&sort=newest}.
    
    \item Innovation Fostering in Accelerator Science and Technology (\acrshort{ifast}) (2021-25)~\cite{IFAST} Horizon 2020 Research and Innovation project working on ``new accelerator designs and concepts, advanced superconducting technologies for magnets and cavities, techniques to increase brightness of synchrotron light sources, strategies and technology to improve energy efficiency, and new societal applications of accelerators'', amongst other areas. WP11 focused on Sustainable Concepts and Technologies (SCAT)~\cite{IFAST_SCAT}.

    \item Accelerator Research and Innovation for European Science and Society (\acrshort{aries}) (2017-21): \url{https://aries.web.cern.ch/}, a project focused on research, development and innovation, with a focus on sustainability through the development of novel concepts, technologies and roadmaps. WP4 was focused on Efficient Energy Management (\acrshort{eem}) (\url{https://www.psi.ch/en/aries-eem}) with tasks working on high efficiency RF power sources, increasing energy efficiency of the spallation target station, high efficiency SRF power conversion and efficient operation of pulsed magnets.

    \item Enhanced European Coordination for Accelerator Research \& Development (\acrshort{eucard-2}) (2014-17) \url{https://eucard2.web.cern.ch/eucard2/}. In particular, WP3 ``Energy Efficiency of Particle Accelerators" (EnEfficient) (\url{https://www.psi.ch/en/enefficient}) developed and promoted concepts for recovering and re-using low-grade waste heat from accelerator facilities. Other work was produced such as on HTS, RF technologies, and novel materials with high-durability and high-thermal-conductivity (which extend component lifetimes).

	\item The STFC Sustainability Principles and Advice for Design and Engineering (\acrshort{spade}) project has contributed many documents and recommendations towards the UK's Greening Government Commitments and \acrshort{stfc}'s net zero by 2050 goal~\cite{UKRIpolicy}. These include studies into sustainable design principles for items like shielding and cryogenics~\cite{spade}. SPADE SharePoint (available to those with a valid UKRI account): \url{https://stfc365.sharepoint.com/sites/SPADE}. 

    \item The \acrshort{erum} Data Hub is a German-based initiative dedicated to the digitization of Fundamental Research in Natural Sciences. The Hub is the central focus point for 20,000 scientists involved in the exploration of the universe and matter (ErUM) across 8 ErUM communities. Its aim is to assist researchers and communities in data management, sustainable software practises, interdisciplinary development and the transfer of digital tools and competencies~\cite{ERUM_sustainability_ethics}. Since August it has launched an Action Group: Sustainability in Digital Transformation focussed on teaching and outreach: \url{https://erumdatahub.de/en/blog/2025/10/07/action-group/}.
\end{itemize}

\subsection{Tools}
\begin{itemize}
    \item openLCA: \url{https://openlca.org/}
    \item One-click LCA: \url{https://oneclicklca.com/}
    \item SparkHub - A new open-access platform for greening research: \url{https://sparkhub.eu/}
    \item \acrshort{ges} Labos 1point5 - a tool for calculating the carbon footprint of a laboratory: \url{https://apps.labos1point5.org/ges-1point5}
    \item Green Usage Impact Logging Tool (\acrshort{guilt}) for SLURM users: \url{https://github.com/GeorgeRoe/guilt}
    \item Review of compute energy evaluation tools: \url{https://github.com/UofM-Green-Compute/Energy-Evaluation-Tools}
    \begin{itemize}
        \item GreenAlgorithms\cite{GreenAlgorithms}: \url{https://www.green-algorithms.org/}
        \item CodeCarbon: \url{https://github.com/mlco2/codecarbon}
        \item Eco2AI~\cite{Eco2AI}: \url{https://github.com/sb-ai-lab/Eco2AI}
        \item CarbonTracker~\cite{CarbonTracker}: \url{https://github.com/lfwa/carbontracker}
        \item MLCO2~\cite{MLCO2}: \url{https://mlco2.github.io/impact/#compute}
        \item CloudCarbonFootprint~\cite{CloudCarbonFootprint}: \url{https://github.com/cloud-carbon-footprint/cloud-carbon-footprint}
        \item Scaphandre: \url{https://github.com/hubblo-org/scaphandre}
        \item Kepler: \url{https://github.com/sustainable-computing-io/kepler}
        \item PowerJoular~\cite{PowerJoularJX}: \url{https://github.com/joular/powerjoular}
        \item JoularJX~\cite{PowerJoularJX}: \url{https://github.com/joular/joularjx}
    \end{itemize}
\end{itemize}
\subsection{Certifications}
\begin{itemize}
    \item ISO 50001~\cite{ISO50001}: \url{https://www.iso.org/iso-50001-energy-management.html}
    \item GreenDiSC~\cite{GreenDiSC}: \url{https://www.software.ac.uk/GreenDiSC}
    \item LEAF~\cite{LEAF}: \url{https://app.ucl.ac.uk/leaf/leaf_external}
    \item MyGreenLab~\cite{MyGreenLab}: \url{https://www.mygreenlab.org/green-lab-certification.html}
\end{itemize}
\subsection{Courses\label{sec:courses}}
\begin{itemize}
    \item CERN lifecycle assessment in design course\footnote{Available to those with a valid CERN account.}:\\ \url{https://lms.cern.ch/ekp/servlet/ekp?PX=N&TEACHREVIEW=N&CID=EKP000044552&TX=FORMAT1&LANGUAGE_TAG=en&DECORATEPAGE=N}.
    \item Creating a Sustainable STFC
    \item The Software Sustainability Institute compiles training, guides and courses relating to all pillars of sustainability, e.g. ``Five recommendations for FAIR software'': \\ \url{https://www.software.ac.uk/training/training-hub}.
\end{itemize}

\subsubsection{Lectures}
    \begin{itemize}
        \item Sustainability of Particle Accelerators at the CERN Accelerator School (\acrshort{cas}) by M. Siedel: \url{https://indico.cern.ch/event/1356988/contributions/5713232/attachments/2936912/5159024/Sustainability_Seidel.pdf}
        \item Environmental Sustainability and Particle Accelerators in the John Adams Institute for Accelerator Science Graduate Accelerator Physics Course by H. Wakeling: \url{https://indico.cern.ch/event/1621143/contributions/6831381/attachments/3229964/5759271/2026_03_04_Lecture_ParticleAcceleratorSustainability.pdf}
    \end{itemize}

\subsection{Workshops and conferences}
\begin{itemize}
	\item Energy for Sustainable Science at Research Infrastructures (\acrshort{essri}) workshop (biannual) \href{https://agenda.ciemat.es/event/4431/}{2026},
    \href{https://agenda.ciemat.es/event/4431/}{2024},
    \href{https://indico.esrf.fr/event/2/}{2022},
    \href{https://indico.psi.ch/event/6754/}{2019} and
    \href{https://indico.eli-np.ro/event/1/}{2017}.
	\item The Sustainable High Energy Physics (Sustainable \acrshort{hep}) workshop (annual) \href{https://indico.global/e/susthep26}{2026},
    \href{https://indico.global/event/4745/}{2025},
    \href{https://indico.global/event/4744/}{2024},
    \href{https://indico.cern.ch/event/1160140/}{2022} and 
    \href{https://indico.cern.ch/event/1004432/}{2021}.
    \item Sustainability Conference for Responsible Research Computing (\acrshort{sc4rc}) (first edition) \href{https://indico.cern.ch/event/1526482/}{2026}.
    \item International Particle Accelerator Conference (\acrshort{ipac}) (annual) Accelerator Technology and Sustainability sessions in \href{https://indico.jacow.org/event/95/}{2026}, \href{https://indico.jacow.org/event/81/}{2025}, \href{https://indico.jacow.org/event/63/}{2024} and \href{https://indico.jacow.org/event/41/}{2023}.
    \item International Conference on High Energy Physics (\acrshort{ichep}) (biannual) Sustainability sessions in \href{https://indico.cern.ch/event/1522800/program}{2026} and \href{https://indico.cern.ch/event/1291157/sessions/543528/#20240719}{2024}.
    \item The Worldwide LHC Computing Grid (\acrshort{wlcg}) Sustainability Forum started in October 2025 and has \href{https://indico.cern.ch/category/20295/}{regular events}.
    \item HEPiX Workshop sustainability tracks in \href{https://indico.cern.ch/event/1598655/program}{Spring 2026}, \href{https://indico.cern.ch/event/1536836/sessions/629614/#20251104.detailed}{Fall 2025}, \href{https://indico.cern.ch/event/1477299/timetable/?view=standard_inline_minutes#b-605424-environmental-sustain}{Spring 2025}, \href{https://indico.cern.ch/event/1450798/sessions/572924/#20241107}{Fall 2024}, and \href{https://indico.cern.ch/event/1377701/sessions/542946/#20240416}{Spring 2024}.
\end{itemize}
\subsubsection{Previously held}
    \begin{itemize}
        \item \href{https://indico.cern.ch/event/1484669/timetable/?view=standard#b-613148-plenary-environmental}{Environmental Sustainability plenary at the WLCG/HSF Workshop 2025}
        \item \href{https://indico.desy.de/event/47133/}{Shaping the Digital Future of ErUM Research: Sustainability \& Ethics 2025}~\cite{ERUM_sustainability_ethics}
        \item \href{https://labos1point5.org/les-colloques/colloque-2024}{Labos 1point5 colloque 2024}
        \item \href{https://tv.theiet.org/Event/Sustainable_Accelerators_Workshop/C763E5C2-6637-4D4C-A4DB-00FFEB7DA7BD}{IET PAEN Sustainable Accelerators Workshop 2024}
        \item \href{https://zenodo.org/records/13309687}{Workshop on efficient magnet- and RF power supplies 2024}
        \item \href{https://indico.cern.ch/event/1376902/}{11th International Workshop on Thin Films
and New Ideas for Pushing the Limits of RF
Superconductivity}
        \item \href{https://indico.desy.de/event/35655/}{Critical Materials and Life Cycle Management: The Example of Rare Earths - curse or blessing? 2023}
        \item \href{https://indico.gsi.de/event/17548/}{Superconductivity for Sustainable Energy Systems and Particle Accelerators 2023}
        \item \href{https://indico.cern.ch/event/1138197/timetable/#all.detailed}{Workshop on efficient RF sources 2022}
    \end{itemize}

\subsection{Suggested publications\label{sec:suggested_publications}}
\begin{itemize}
    \item ``Sustainability Considerations for Accelerator and Collider Facilities'' from the ICFA Panel for Sustainable Accelerators and Colliders~\cite{SustainabilityConsiderations}
    \item ``Sustainability and Carbon Emissions of Future Accelerators'' by K. Bloom and V. Boisvert (2025)~\cite{CO2-future-accelerators}.
    \item ``Sustainability Assessment of Future Accelerators'' from the Laboratory Directors Group Sustainability Working Group (ESPPU 2025 input)~\cite{LDG-ESPPU}
    \item ``Input to European Strategy Update for Particle Physics: Sustainability'' (ESPPU 2025 input)~\cite{independent-ESPPU}
    \item ``Countering the biodiversity loss using particle physics research sites'' (ESPPU 2025 input)~\cite{biodiversity-ESPPU}
    \item ``Ten simple rules to make your computing more environmentally sustainable"~\cite{BPcomputing10}.
    \item ``Energy efficiency trends in \acrshort{hpc}: what high-energy and astrophysicists need to know''~\cite{HPC_energy_eff}
    \item ``Resource-aware Research on Universe and Matter: Call-to-Action in Digital Transformation''~\cite{call-to-action}
    \item ``Towards Climate Sustainability of the Academic System in Europe and beyond''~\cite{allea_academic_system}
\end{itemize}

\subsubsection*{LCAs or environmental assessments for research facilities \label{sec:LCA_publications}}
\begin{itemize}
    \item Future Circular Collider Feasibility Study Report : Volume 3 Civil Engineering, Implementation and Sustainability (2025)~\cite{FCC_sustainability}: \\ \url{https://cds.cern.ch/record/2928194?ln=en}
    \item Whole Lifecycle Assessment of CLIC and ILC - Phase 2 (2025)~\cite{ARUPstudyCLICILC}: \\ \url{https://edms.cern.ch/document/3283864/1}
    \item Sustainability for Particle Accelerators: RUEDI - A Case Study (2024)~\cite{ruedi}: \\ \url{https://www.astec.stfc.ac.uk/news/sustainability-for-particle-accelerators-ruedi-a-case-study/}
    \item Sustainable astronomy: A comparative life cycle assessment of off-grid hybrid energy systems to supply large telescopes (2024)~\cite{SustAstroOffGrid}: \\ \url{https://link.springer.com/article/10.1007/s11367-024-02288-9}
    \item FCC Construction Carbon Footprint Benchmark and Optimisation Strategies (2024)~\cite{FCC_carbon_optimisation}: \url{https://zenodo.org/records/13899160}
    \item Assessment of the environmental impacts of the Cherenkov Telescope Array Mid-Sized Telescope (2024)~\cite{CherenkovTele}: \url{https://arxiv.org/abs/2406.17589}
    \item Life Cycle Assessment of the Athena X-ray Integral Field Unit (2024)~\cite{AthenaTele}: \\ \url{https://arxiv.org/abs/2404.15122}
    \item CLIC and ILC Life Cycle Assessment Final Report - Phase 1 (2023)~\cite{ARUPstudyCLICILC}: \\ 
    \url{https://edms.cern.ch/document/2917948/1}
    \item Sustainability Strategy for the Cool Copper Collider (2023)~\cite{C3}: \\ \url{https://journals.aps.org/prxenergy/pdf/10.1103/PRXEnergy.2.047001}
    \item Life cycle analysis of the GRAND experiment (2023)~\cite{GRANDlca}: \\ \url{https://arxiv.org/abs/2309.12282}
    \item Estimating the carbon footprint of the GRAND experiment (2021)~\cite{GRANDcarbonfootprint}: \\ \url{https://arxiv.org/abs/2101.02049}
\end{itemize}

%%%%%%%%%%%%%%%%%%%%%%%%%%%%%%%%%%%%%%%%%%%%%%%%%%%%%%%%
\section*{Future directions for this living document}

This section briefly outlines the authors' considerations for the next stages of this living document. 

First, to enhance its usefulness as a resource, the authors are exploring implementation of a structured recommendation framework that communicates the anticipated impact of each recommendation (high/medium/low), and potentially the difficulty of implementation, for example by referencing Technology Readiness Level where appropriate.
Second, the authors aim to incorporate more quantitative analyses, including target figures for selected recommendations based on relevant case studies that have implemented these measures.
Third, the authors aim to highlight key gaps in tools and knowledge, further than already included in the document. One example is the need for tools to further the ease of implementation of lifecycle emissions evaluation efforts. One recently published example in another field is the Open-source Rocket and Constellation Lifecycle Emissions (ORACLE) GitHub repository~\cite{ORACLE_rocket}. By making their code and data accessible, the authors of the ORACLE project wish to encourage transparency and to foster collaboration within space sustainability. The authors of this living document applaud efforts like this, and wish to encourage and promote similar efforts within fields applicable to accelerator facilities, and ultimately research overall.
Fourth, the authors endeavour to continuously add and update evidence, case studies and context to the document as and when available. They aim to publish updates to this document when sufficient amendments have been made. What constitutes sufficient amendments could include anything from minor additions or updates, to a major amendment the authors deem could have significant impact on sustainability.

\backmatter

\bmhead{Supplementary information}

Not applicable.

\bmhead{Acknowledgements}

The work of lead/corresponding author Hannah Wakeling was supported by the STFC via the John Adams Institute for Accelerator Science (JAI), University
of Oxford [Grant No. ST/V001655/1].

The authors would like to acknowledge and thank those that worked on the case studies and resources referenced within this document. Without these fantastic and ongoing efforts, this document would not be the same. 
In addition, the authors would like to thank the individuals who provided feedback on the previous version of this living document. Your insight helped shape this version to its current form. 

\section*{Declarations}

\begin{itemize}
    \item Author Ben Shepherd has worked with ZEPTO and is co-author of the ``Sustainability for Particle Accelerators: RUEDI - A Case Study'' and the ``Sustainability Assessment of Future Accelerators'' from the Laboratory Directors Group Sustainability Working Group.
    \item Author Niko Neufeld is co-author of the ``Sustainability Assessment of Future Accelerators'' from the Laboratory Directors Group Sustainability Working Group.
    \item Author Dwayne Spiteri is funded by the \acrshort{rf2.0} project that has received funding from the European Union’s Horizon Europe research and innovation programme under grant agreement No. 101131850 and from the Swiss State Secretariat for Education Research and Innovation (\acrshort{seri}). DS is a member of the ErUM data-hub sustainability action group and an author on the paper ``Shaping the Digital Future of ErUM Research: Sustainability \& Ethics'', a research co-ordinator for the SUSFECIT project, and a session chair for environmental sustainability, business continuity and facility improvement for HEPiX 2026.
    \item Author Jim Clarke is a member of lab directors group and the European strategy group for the current \acrshort{esppu}.
    \item Author John Thomason is project sponsor and Senior Responsible Owner for the ISIS-II Project.
    \item Author Hannah Wakeling is Sustainability Lead for the ISIS-II Project and author of the cited ``Updated results of a life cycle assessment of the ISIS-II neutron and muon source''. HW is also co-organiser of the Sustainable \acrshort{hep} conference 2024, 2025 and 2026. HW is co-author of mentioned publications: ``Input to European Strategy Update for Particle Physics: Sustainability'', and the ``Sustainability Assessment of Future Accelerators'' from the Laboratory Directors Group Sustainability Working Group. HW is author of the mentioned ``Environmental Sustainability and Particle Accelerators lecture'' in the John Adams Institute for Accelerator Science Graduate Accelerator Physics Course. 
\end{itemize}

\begin{appendices}

\printglossary[type=\acronymtype]

\end{appendices}

\bibliography{sn-bibliography}

\end{document}